\documentclass[preprint,nofootinbib,floatfix,a4paper,prd,superscriptaddress]{revtex4-1}
\usepackage{hyperref}
\usepackage{graphicx}
\usepackage[utf8]{inputenc} 
\usepackage{amsmath,amssymb}
\usepackage{slashed}
\usepackage{epstopdf}
\usepackage{geometry}

\usepackage{rotating}

\usepackage{tabularx}

\newcolumntype{C}{>{\centering\arraybackslash}X} 
\usepackage[dvipsnames]{xcolor}
\def\be{\begin{equation}}
\def\ee{\end{equation}}
\def\bea{\begin{eqnarray}}
\def\eea{\end{eqnarray}}

\newcommand{\oh}{\ensuremath{\Omega_{D}h^2}}

\newcommand{\sigsip}{\ensuremath{\sigma^{\rm{SI}}_p}}

\newcommand{\gev}{\ensuremath{\text{~GeV}}}

\def  \bcen   {\begin{center}}
\def  \ecen   {\end{center}}
\def  \beq    {\begin{equation}}
\def  \eeq    {\end{equation}}
\def  \bpm    {\begin{pmatrix}}
\def  \epm    {\end{pmatrix}}
\def  \beqa   {\begin{eqnarray}}
\def  \eeqa   {\end{eqnarray}}
\def\bea{\begin{eqnarray}}
\def\eea{\end{eqnarray}}
\def  \nn     {\nonumber }

\def\la   {\lambda}

\def\nn{\nonumber}
\def\lee { \left( }
\def\rii { \right) }
\def\lan   {\langle}
\def\ran   {\rangle}

\def\De {\Delta}

\def\to {\rightarrow}

\begin{document}

\title{Effects of New Heavy Fermions on Complex Scalar Dark Matter Phenomenology in Gauged Two Higgs Doublet Model}

\author{Bayu Dirgantara}
\email{bayuquarkquantum@yahoo.com}
\affiliation{\small School of Physics and Center of Excellence in High Energy Physics $\textit{\&}$ Astrophysics, Suranaree University of Technology, Nakhon Ratchasima, 30000, Thailand}

\author{Chrisna Setyo Nugroho}
\email{csnugroho@cts.nthu.edu.tw}
\affiliation{\small Physics Division, National Center for Theoretical Sciences, National Tsing-Hua University, Hsinchu, 30013, Taiwan}


\preprint{NCTS-PH/2012}

\begin{abstract}

We study the inclusion of new heavy fermions on complex scalar dark
matter (DM) phenomenology within gauged two Higgs doublet model
(G2HDM). We find that for DM mass above 1 TeV, heavy quarks
coannihilations into the Standard Model (SM) quarks and gluons
dominate the thermally-averaged cross section relevant for the relic
abundance of complex scalar DM. We demonstrate that the effects of
QCD Sommerfeld correction as well as QCD bound state formation in
determining the DM relic density are negligible. We show that the allowed
parameter space is significantly constrained by the current PLANCK
relic density data as well as XENON1T limit appropriate for DM
direct search.  
 
\end{abstract}

\maketitle

\section{Introduction}

The nature of dark matter (DM) is one of the open problems in cosmology, astrophysics and particle physics.
Apart from its gravitational interaction,
there is no evidence that it interacts with ordinary matter
via the other existing forces in nature.
Furthermore, the current understanding of our universe
can be explained very well if one includes the existence of the cold
DM in addition to the ordinary matter and dark energy.
It has been assumed that the DM was in thermal equilibrium with 
the Standard Model (SM) particles in the early universe. As the
reaction rate that keeps the DM in thermal equilibrium with other SM
particles drops out and becomes comparable with the expansion rate of the
universe, it froze-out. As a result, it remains with us today.
One of the most popular and well studied DM candidates so far is  
the weakly interacting massive particle (WIMP). This candidate fits the observed DM abundance with the typical electroweak scale annihilation cross section.
 
The need to extend the SM particle contents is necessary since 
it does not provide any suitable DM candidate. 
One of the most studied examples of the SM extension is the general two
Higgs doublet model (2HDM).
This model comes with many variations and each of them offers  different
phenomenological features,
see for example~\cite{Branco:2011iw,Logan:2014jla} for a review.
As one of the 2HDM variants, the inert Higgs doublet model (IHDM)~\cite{Deshpande:1977rw}
provides a suitable DM candidate. This is done by imposing the
discrete $\mathcal Z_2$ symmetry in which the second Higgs doublet
belongs to the $\mathcal Z_2$-odd particle.
This model has been studied in detail over the years~\cite{LopezHonorez:2006gr, Arina:2009um, Nezri:2009jd, Miao:2010rg,
Gustafsson:2012aj, Arhrib:2012ia, Arhrib:2014pva,Swiezewska:2012eh, Arhrib:2013ela,
Goudelis:2013uca, Krawczyk:2013jta, Krawczyk:2013pea, Ilnicka:2015jba,
Diaz:2015pyv, Modak:2015uda, Kephart:2015oaa,Queiroz:2015utg, Garcia-Cely:2015khw,
Hashemi:2016wup, Poulose:2016lvz, Alves:2016bib, Datta:2016nfz,
Belyaev:2016lok, Belyaev:2018ext}. 
Recently, this ${\mathcal Z}_2$ symmetry appears as an accidental symmetry 
in a renormalizable gauged two Higgs doublet model
(G2HDM) and it has been studied in detail in~\cite{Huang:2015wts,Chen:2019pnt}. 
In this model, the two Higgs doublets are put together in a doublet 
representation of an additional $SU(2)_H$ gauge group. 
In addition to this, there is also a new $U(1)_X$ symmetry. As for the scalar sector, 
it is expanded by including a new $SU(2)_H$ triplet and doublet. These new scalars transform trivially under the SM gauge group.

As in IHDM, there is a suitable scalar DM candidate in G2HDM. In this case, the stability of the DM is protected by the accidental
$\mathcal Z_2$ symmetry. The phenomenological study of this DM has
been done in~\cite{Chen:2019pnt}. This study has taken the previous
results of the G2HDM studies~\cite{Huang:2015rkj, Huang:2017bto,
Chen:2018wjl, Arhrib:2018sbz, Huang:2019obt} into account, especially the results from~\cite{Arhrib:2018sbz, Huang:2019obt}.
These two studies put the constraints in the scalar sector and gauge sector of the G2HDM.
These include vacuum stability of the scalar potential, perturbative unitarity, 
Higgs physics, Drell-Yan process, the $Z^{'}$ search, as well as
electroweak precision test (EWPT). We dubbed these results as the
scalar and gauge sector constraints ({\bf SGSC}). However, the
previous DM study in~\cite{Chen:2019pnt} neglects the heavy $\mathcal Z_2$-odd fermions in their calculation. In this paper, 
we focus on the effects of these new heavy fermions on
complex scalar DM phenomenology. For consistency with the previous DM study, we take {\bf SGSC} constraints as our starting points.

This paper is organized as follows.
In Sec.~\ref{sec:model} we briefly mention the important aspects of the G2HDM
model, especially the scalar potential, mass spectra, and suitable dark matter candidates.
To compare our DM study against the existing experimental data, we discuss
how relic density ({\bf RD}) and direct detection ({\bf
DD}) constraint our complex scalar DM. This is discussed in Sec.~\ref{sec:constraints}.  
In Sec.~\ref{sec:result}, we briefly explain the methodology employed in our numerical
computation. We further discuss in detail the results obtained in our analysis.
In Sec.~\ref{sec:SCBSF}, we discuss QCD Sommerfeld correction and QCD bound state effect relevant for the relic density calculation.
The allowed parameter space after imposing DM physics constraints
is discussed in Sec.~\ref{sec:ps}.
Finally, we summarize and conclude our study in Sec.~\ref{sec:summary}. 
We list the most important Feynman rules used in the discussion 
of this work in Appendix~\ref{sec:appendix}.

\vfill

\section{The G2HDM Model}
\label{sec:model}

\subsection{Matter Content} 
\begin{table}[htbp!]
	\begin{tabular}{|c|c|}
		\hline
		Matter Fields & $SU(3)_C \times SU(2)_L \times SU(2)_H \times U(1)_Y \times U(1)_X$ \\
		\hline \hline
		$H=\left( H_1\,, \; H_2 \right)^{\rm T}$ & (1, 2, 2, 1/2, 1) \\
		$\Delta_H$ & (1, 1, 3, 0, 0) \\
		$\Phi_H$ & (1, 1, 2, 0, $1$) \\
		\hline\hline
		$Q_L=\left( u_L\,, \; d_L \right)^{\rm T}$ & (3, 2, 1, 1/6, 0)\\
		$U_R=\left( u_R\,, \; u^H_R \right)^{\rm T}$ & (3, 1, 2, 2/3, 1) \\
		$D_R=\left( d^H_R\,, \; d_R \right)^{\rm T}$ & (3, 1, 2, $-1/3$, $-1$) \\
		\hline
		$L_L=\left( \nu_L\,, \; e_L \right)^{\rm T}$ & (1, 2, 1, $-1/2$, 0) \\
		$N_R=\left( \nu_R\,, \; \nu^H_R \right)^{\rm T}$ & (1, 1, 2, 0, $1$) \\
		$E_R=\left( e^H_R\,, \; e_R \right)^{\rm T}$ & (1, 1, 2, $-1$, $-1$) \\
		\hline
		$\nu_L^H$ & (1, 1, 1, 0, 0) \\
		$e_L^H$ & (1, 1, 1, $-1$, 0) \\
		\hline
		$u_L^H$ & (3, 1, 1, 2/3, 0) \\
		$d_L^H$ & (3, 1, 1, $-1/3$, 0) \\
		\hline
	\end{tabular}
	\caption{Matter contents and their corresponding quantum numbers in G2HDM. New heavy fermions are denoted by the superscript $H$.}
	\label{tab:quantumnos}
\end{table}
The gauge group of G2HDM is realized by expanding
the SM gauge symmetry with additional $SU(2)_H \times U(1)_X$ dubbed
as hidden gauge sector. The SM scalar sector $H_{1}$ is extended by
adding one Higgs doublet $H_{2}$ in such a way that both of them
transform under the fundamental representation of 
$SU(2)_L$ and $SU(2)_H$ gauge group.
Furthermore, $\Delta_H$ and $\Phi_H$, which transform under
the triplet and doublet representations of $SU(2)_H$, are present. Both of them are SM singlets.

New
right-handed heavy fermions are put together with the SM
right-handed fermions into $SU(2)_H$ doublets while maintaining
the trivial representation of $SU(2)_L$.
Furthermore, anomaly cancellation dictates us to further include
two pairs of left-handed heavy quarks and two pairs of left-handed heavy
leptons for each generation, which are singlets under $SU(2)$ and
$U(1)_X$.
As a remark, 
one notes that the $SU(2)_H$ considered here is different from the $SU(2)_R$ in 
left-right symmetric models~\cite{Mohapatra:1979ia,Keung:1983uu}.
The $W^{\prime (p,m)}$ in G2HDM is electrically neutral, while the $W^{\prime \pm}$ in 
left-right symmetric carry non-zero electric charges. This is the rationale behind the superscripts $p$ and $m$ in labelling $W^{\prime}$.
As another comparison, we also observe that non-sterile right-handed
neutrinos $\nu_{l R}$s, incorporated in the mirror fermion models of
electroweak scale right-handed neutrinos~\cite{Hung:2006ap,Hung:2017voe,Chang:2016ave,Hung:2015hra,Chang:2017vzi,Hung:2017tts}, 
are different from our setup.
In the mirror fermion models, $\nu_{l R}$s are paired together with mirror charged leptons $l^M_R$s to form $SU(2)_L$ doublets.
In contrast, they are grouped with new heavy right-handed
neutrinos $\nu^H_{l R}$ 
to form $SU(2)_H$ doublets in G2HDM.
There are several models that impose 
additional gauge symmetry on 2HDM to solve flavor problem, 
dark matter and neutrino masses, see for instance~\cite{Ko:2012hd,Campos:2017dgc,Camargo:2018klg,Camargo:2018uzw,Camargo:2019ukv,Cogollo:2019mbd}.
In table~\ref{tab:quantumnos}, we list the matter contents
of the G2HDM model and their associated quantum numbers.

%

\subsection{Scalar Potential and Mass Spectra}
\label{subsec:pottheorcons}

\subsubsection*{Scalar Potential}
The most general renormalizable scalar potential that satisfies the G2HDM symmetries comprises 4 distinct terms ~\cite{Arhrib:2018sbz}
\begin{equation}
V_T = V (H) + V (\Phi_H ) + V ( \De_H ) + V_{\rm mix} \left( H, \Delta_H, \Phi_H \right) \; .
\label{eq:higgs_pots} 
\end{equation}
The first term $V(H)$ in Eq.~\eqref{eq:higgs_pots} contains self-interaction of $SU(2)_L$ and $SU(2)_H$ scalar doublet $H$ reads  as
\begin{align}
\label{VH1H2}
V(H)
= {}& \mu^2_H   \left( H^{\alpha i}  H_{\alpha i} \right)
+  \la_H \left( H^{\alpha i}  H_{\alpha i} \right)^2  
+ \frac{1}{2} \la'_H \epsilon_{\alpha \beta} \epsilon^{\gamma \delta}
\left( H^{ \alpha i}  H_{\gamma  i} \right)  \left( H^{ \beta j}  H_{\delta j} \right) \nn \\
= {}& \mu^2_H   \left( H^\dag_1 H_1 + H^\dag_2 H_2 \right) 
+ \la_H   \left( H^\dag_1 H_1 + H^\dag_2 H_2 \right)^2  \nn \\
{}& \;\;\;\;\;\;\;\;\;\;\;\;\;\;\;\;\;\;\;\;\;\;\;\;\;\;\;\;\;\;\; + \la'_H \left( - H^\dag_1 H_1 H^\dag_2 H_2 
+ H^\dag_1 H_2 H^\dag_2 H_1 \right)  \; , 
\end{align}
where Greek and Latin letters indicate $SU(2)_H$ and $SU(2)_L$
indices respectively, 
both of which run from 1 to 2, and the upper and lower indices are
related by complex conjugation, i.e., $H^{\alpha i} = H^*_{\alpha i}$. A close inspection shows that the second line of Eq.~\eqref{VH1H2} exhibits the discrete $\mathcal Z_2$ symmetry of $H_1 \to H_1$ and $H_2 \to - H_2$. The appearance of this discrete symmetry in G2HDM model is more natural unlike the discrete symmetry
in general 2HDM model. In the latter case, one needs to put this 
symmetry by hand to forbid FCNC at tree level in the Yukawa sectors.
The second term $V ( \Phi_H )$ is the self interaction of $\Phi_H$ 
\begin{align}
\label{VPhi}
V ( \Phi_H )
= {}& \mu^2_{\Phi}   \Phi_H^\dag \Phi_H  + \la_\Phi \lee \Phi_H^\dag \Phi_H  \rii^2 \nn \\
= {}& \mu^2_{\Phi} \lee \Phi^*_1\Phi_1 + \Phi^*_2\Phi_2 \rii +  \la_\Phi \lee \Phi^*_1\Phi_1 + \Phi^*_2\Phi_2 \rii^2 \; , 
\end{align}
where $\Phi_H = (\Phi_1\,, \; \Phi_2)^{\rm T}$ belongs to the $SU(2)_H$ doublet. The self interaction of the $SU(2)_H$ scalar triplet $\Delta_{H}$ reads
\begin{align}
\label{VDeltas}
V ( \De_H ) = {}& - \mu^2_{\De} {\rm Tr} \lee \De^2_H  \rii  \;  + \la_\De \lee {\rm Tr} \lee \De^2_H  \rii \rii^2 \nn \\
= {}& - \mu^2_{\De} \lee \frac{1}{2} \De^2_3 + \De_p \De_m  \rii +  \la_{\De} \lee \frac{1}{2} \De^2_3 + \De_p \De_m  \rii^2 \; , 
\end{align}
where the triplet field is expressed as 
\begin{align}
\De_H=
\begin{pmatrix}
	\De_3/2   &  \De_p / \sqrt{2}  \\
	\De_m / \sqrt{2} & - \De_3/2   \\
\end{pmatrix} = \De_H^\dagger \; {\rm with}
\;\; \Delta_m = \left( \Delta_p \right)^* \; {\rm and} \; \left( \Delta_3 \right)^* = \Delta_3 \;   .
\end{align}
Here, the off-diagonal components of the 
$SU(2)_H$ triplet, $\Delta_{H}$, are electrically neutral. 
Following the same labelling as in $W^{\prime (p,m)}$, 
we put the subscripts $p$ and $m$ on them. Finally, the last term $V_{\rm{mix}}$ takes all possible mixing between $H$, $\Phi_H$ as well as $\De_H$ into account, and it is given by
\begin{align}
\label{VvMix}
V_{\rm{mix}} \left( H , \Delta_H, \Phi_H \right) = 
{}& + M_{H\De}  \lee H^\dag \De_H H \rii -  M_{\Phi\De}  \lee \Phi_H^\dag \De_H \Phi_H \rii  \nn \\
{}& + \la_{H\Phi} \lee H^\dag H  \rii  \lee \Phi_H^\dag \Phi_H \rii  
+ \la^\prime_{H\Phi} \lee H^\dag \Phi_H  \rii  \lee \Phi_H^\dag H \rii
\nn\\
{}& +  \la_{H\De} \lee H^\dag H  \rii    {\rm Tr} \lee \De^2_H  \rii
+ \la_{\Phi\De} \lee \Phi_H^\dag \Phi_H \rii {\rm Tr} \lee \De^2_H \rii  \; . 
\end{align}
The explicit expression of Eq.~\eqref{VvMix} in terms of its components is  
\begin{align}
V_{\rm{mix}} \left( H , \Delta_H, \Phi_H \right) =
{}& + M_{H\De} \lee \frac{1}{\sqrt{2}}H^\dag_1 H_2 \De_p  
+  \frac{1}{2} H^\dag_1 H_1\De_3 + \frac{1}{\sqrt{2}}  H^\dag_2 H_1 \De_m  
- \frac{1}{2} H^\dag_2 H_2 \De_3   \rii   \nn \\
{}& - M_{\Phi\De} \lee  \frac{1}{\sqrt{2}} \Phi^*_1 \Phi_2 \De_p  
+  \frac{1}{2} \Phi^*_1 \Phi_1\De_3 + \frac{1}{\sqrt{2}} \Phi^*_2 \Phi_1 \De_m  
- \frac{1}{2} \Phi^*_2 \Phi_2 \De_3   \rii  \nn \\
{}& +  \la_{H\Phi} \lee H^\dag_1 H_1 + H^\dag_2 H_2 \rii  \lee \Phi^*_1\Phi_1 + \Phi^*_2\Phi_2 \rii \nn\\
{}& +  \la^\prime_{H\Phi} \lee H^\dag_1 H_1 \Phi^*_1\Phi_1 + H^\dag_2 H_2  \Phi^*_2\Phi_2 
+ H^\dag_1 H_2 \Phi_2^*\Phi_1 + H^\dag_2 H_1  \Phi^*_1\Phi_2  \rii \nn\\
{}& + \la_{H\De} \lee H^\dag_1 H_1 + H^\dag_2 H_2 \rii   \lee \frac{1}{2} \De^2_3 + \De_p \De_m  \rii \nn\\
{}& + \la_{\Phi\De}  
\lee  \Phi^*_1\Phi_1 + \Phi^*_2\Phi_2 \rii  \lee \frac{1}{2} \De^2_3 + \De_p \De_m  \rii \; .
\label{eq:vmix}
\end{align}
As before, $V_{\rm{mix}} \left( H , \Delta_H, \Phi_H \right)$ is also invariant under $H_1 \to H_1$, $H_2 \to - H_2$, $\Phi_1 \to - \Phi_1$, $\Phi_2 \to  \Phi_2$, $\De_3 \to  \De_3$, and $\De_{p,m} \to  - \De_{p,m}$. It has been shown in~\cite{Chen:2019pnt} that this discrete symmetry holds in all sector in G2HDM. 

\subsection{Mass Spectra}
\label{sec:MassSpectrum}
The gauge symmetry of G2HDM is broken spontaneously by the vacuum expectation values (VEVs) 
of $\langle H_1 \rangle = (0, v/\sqrt 2)^{\rm T}$, 
$\langle \Phi_2 \rangle = v_\Phi/\sqrt 2$, and $\lan \De_3 \ran = - v_\De$~\cite{Huang:2015wts,Arhrib:2018sbz}. As a result of this spontaneous symmetry breaking (SSB), all fields in G2HDM acquire their mass from these VEVs. In this subsection, we discuss the mass spectra of
G2HDM,
focusing mainly on the scalar and gauge sector. 

\subsubsection*{Higgs-like  ($\mathcal Z_2$-even) Scalars}

The mass terms as well as the mixing terms of the
scalar fields can be extracted by writing the scalar potential in terms
of the existing VEVs and further taking the second derivatives with
respect to the corresponding scalar fields. The SM Higgs can be obtained from the mixing of
three real scalars $h$, $\phi_2$ and $\delta_3$~\footnote{We parameterize the scalar fields as the notations of~\cite{Huang:2015wts}:
$H_1 = \left( \begin{array}{c} G^+ \\ \frac{v + h}{\sqrt 2} + i G^0 \end{array} \right)$, 
$H_2 = \left( \begin{array}{c} H^+ \\ H^0_2 \end{array} \right)$, 
$\Phi_H = \left( \begin{array}{c} G^p_H \\ \frac{v_\Phi + \phi_2}{\sqrt 2} + i G^0_H \end{array} \right)$, 
and 
$\De_H = \left( 
\begin{array}{cc} 
\frac{-v_\De + \delta_3}{2} & \frac{\De_p}{\sqrt 2} \\ \frac{\De_m}{\sqrt2} & \frac{v_\De - \delta_3}{2}
\end{array} \right)$.}. 
In the basis of ${\cal S}=\{h, \phi_2, \delta_3\}^{\rm T}$, the corresponding mixing matrix of these ${\cal Z}_2$-even neutral real scalars is 
\begin{align}
{\mathcal M}_0^2 =
\begin{pmatrix}
	2 \lambda_H v^2 & \lambda_{H\Phi} v v_\Phi 
	& \frac{v}{2} \left( M_{H\De} - 2 \lambda_{H \De} v_\De \right)  \\
	\lambda_{H\Phi} v v_\Phi
	& 
	2 \lambda_\Phi v_\Phi^2
	&  \frac{ v_\Phi}{2} \left( M_{\Phi\De} - 2 \lambda_{\Phi \De} v_\De \right) \\
	\frac{v}{2} \left( M_{H\De} - 2 \lambda_{H \De} v_\De \right)  & \frac{ v_\Phi}{2} \left( M_{\Phi\De} - 2 \lambda_{\Phi \De} v_\De \right) & \frac{1}{4 v_\De} \left( 8 \lambda_\De v_\De^3 + M_{H\Delta} v^2 + M_{\Phi \De} v_\Phi^2 \right)   
\end{pmatrix} \; .
\label{eq:scalarbosonmassmatrix}
\end{align}
This matrix is diagonalized by using an orthogonal matrix ${\cal O}$ to obtain the  masses of the physical fields as
\begin{equation}
{\cal O}^{\rm T}\cdot {\mathcal M}_0^2 \cdot {\cal O} = {\rm Diag}(m^2_{h_1}, m^2_{h_2}, m^2_{h_3}) \; ,
\label{eq:OTM0sqO}
\end{equation}
where the masses of the fields are ordered as $m_{h_1} \leq
m_{h_2} \leq m_{h_3}$. 
The interaction basis ${\cal S}$ and the mass eigenfields are linked
to each other via ${\cal O}$ as ${\cal S} = {\cal O}\cdot\{h_1, h_2, h_3\}^{\rm T}$. The 125 GeV SM Higgs is chosen to be the
lightest mass eigenstate $h_{1}$.

The remaining ${\cal Z}_2$-even scalars $G^{\pm,0}$ and $G^0_H$ do not mix with the existing scalar fields. These would be Goldstone bosons will be absorbed by the gauge bosons.

\subsubsection*{Dark  ($\mathcal Z_2$-odd) Scalars}

The mass of the charged Higgs can be read directly from the scalar potential since it does not mix with other scalars in G2HDM. Since it interacts with $H_1$, $\Phi_H$ as well as $\Delta_H$, the charged Higgs receives its mass from the VEVs of these fields as
\begin{align} m^2_{H^\pm} &= M_{H \De} v_\De  - \frac{1}{2}\la^\prime_H v^2
+\frac{1}{2}\lambda^\prime_{H\Phi}v_\Phi^2 \;.  \label{chargedHiggsmass}
\end{align}

On the other hand, the three neutral complex fields $G^{p,m}_H$ , $H^{0(*)}_2$ and $\Delta_{p,m}$~\footnote{See the previous footnote for the definitions of these complex scalars.} mix with each other as 
\begin{align}
{\mathcal M}_0^{\prime 2} =
\begin{pmatrix}
	M_{\Phi \Delta} v_\Delta +\frac{1}{2}\lambda^\prime_{H\Phi}v^2 & \frac{1}{2}\lambda^\prime_{H\Phi}  v v_\Phi & - \frac{1}{2} M_{\Phi \Delta} v_\Phi  \\
	\frac{1}{2}\lambda^\prime_{H\Phi} v v_\Phi &  M_{H \Delta} v_\Delta
	+\frac{1}{2}\lambda^\prime_{H\Phi} v_\Phi^2
	&  
	\frac{1}{2} M_{H \Delta} v\\
	- \frac{1}{2} M_{\Phi \Delta} v_\Phi & \frac{1}{2} M_{H \Delta} v & 
\frac{1}{4 v_\Delta} \left( M_{H\Delta} v^2 + M_{\Phi \Delta} v_\Phi^2 \right)\end{pmatrix} \, ,
\label{eq:Z2oddmassmatrix}
\end{align}
where this matrix is written in the basis of $\mathcal{G}=\{ G^p_H , H^{0*}_2, \Delta_p  \}^T$. As one can easily check from its
determinant, this mass matrix has at least one zero eigenvalue. Upon
diagonalization, the interaction basis and mass basis are related via
orthogonal matrix $\mathcal{O}^D$ as $\mathcal{G} = {\cal O}^D\cdot \{ \widetilde{G}^p, D, \widetilde{\Delta}\}^{\rm T}$ such that the
orthogonal matrix $\mathcal{O}^D$ brings the mixing matrix into the
diagonal form 
\begin{equation}
({\cal O}^D)^{\rm T}\cdot {\mathcal M}_0^{\prime 2} \cdot {\cal O}^D =
{\rm Diag}(0, m^2_D, m^2_{\widetilde\Delta}) \; .
\label{eq:OTM0sqO2}
\end{equation}
The complex gauge bosons $SU(2)_H$, $W^{\prime \, (p,m)}$, become massive by absorbing the first zero eigenvalue in Eq.~\eqref{eq:OTM0sqO2}, i.e., $\widetilde{G}^{p,m}$, into their longitudinal component. Here, the hierarchy $m^2_D < m^2_{\widetilde{\Delta}}$ has been assumed. In order to facilitate the analysis in the case when $D$ is the dark matter candidate, we have excluded $m_{D}=m_{\widetilde{\Delta}}$ case. This degenerate in masses happens actually under the very special condition as can be inferred from the mass formula of the two physical states given below
\begin{equation}
\label{darkmattermass}
M^2_{D, {\widetilde \Delta}} = \frac{-B \mp \sqrt{B^2 - 4 A C}}{2A} \; , 
\end{equation}
where
\begin{align}
\label{ABC}
A &= 8 v_\Delta \; , \nonumber \\ 
B & = - 2 \left[ M_{H\Delta} \left( v^2 + 4 v_\Delta^2 \right) + M_{\Phi \Delta}
	\left( 4 v_\Delta^2 + v_\Phi^2 \right) + 2 \lambda^\prime_{H\Phi} v_\Delta
\left( v^2 + v_\Phi^2 \right) \right] \;, \\
C & = \left( v^2 + v_\Phi^2 + 4
	v_\Delta^2 \right) \left[ M_{H \Delta} \left( \lambda^\prime_{H\Phi} v^2 + 2
	M_{\Phi \Delta} v_\Delta \right) + \lambda^\prime_{H\Phi} M_{\Phi \Delta}
v_\Phi^2  \right] \; .\nonumber
\end{align}

As can be seen from this formula, the degeneracy between $D$ and $\widetilde{\Delta}$ occurs when the argument under the square root vanishes. Furthermore, we require $m_{H^\pm}>m_D$ to make $D$ a suitable DM candidate.

\subsubsection*{Gauge Bosons}

Once the SSB takes place, all the gauge fields $W$, $W'$, $B$, and $X$ become massive. The charged SM gauge boson, $W^\pm$, doesn't mix
with other gauge bosons. Its mass is completely determined by the
VEV of $H_{1}$ as $m_{W^{\pm}} = gv/2$. On the other hand, the neutral gauge
boson coming from the off-diagonal component of $SU(2)_H$, $W^{\prime (p,m)}$, receives its mass from all existing VEVs as
\begin{equation}
m^2_{W^{\prime (p,m)}}  = \frac{1}{4} g^2_H \lee v^2 + v^2_\Phi + 4 v^2_\De \rii.  \; 
\label{eq:Wppmmass}
\end{equation}
The neutral gauge bosons corresponding to the third generator of the $SU(2)_L$ and $SU(2)_H$, $W^3$ and $W^{\prime 3}$, mix with $B$ as well as $X$. In the basis of ${\mathcal V}^\prime = \{B,W^{3},W^{\prime 3},X\}^{\rm T}$, their mixing matrix is given by the following 4$\times$4 matrix
\begin{equation}
\label{M1sq2}
{\mathcal M}_1^2 =     \begin{pmatrix}
	\frac{g^{\prime 2} v^2 }{4} + M_Y^2 & - \frac{g^{\prime} g \, v^2 }{4}  &  \frac{g^{\prime} g_H v^2 }{4} & \frac{g^\prime g_X v^2}{2} + M_X M_Y \\
	- \frac{g^{\prime} g \, v^2 }{4} & \frac{ g^2 v^2 }{4} & - \frac{g g_H v^2 }{4} 
	& - \frac{ g g_X v^2  }{2} \\
	\frac{g^{\prime} g_H v^2 }{4} & - \frac{g g_H v^2 }{4}   & \frac{g^2_H  \lee v^2 + v^2_\Phi \rii }{4}  & 
	\frac{g_H g_X \lee v^2 - v^2_\Phi \rii }{2} \\ 
	\frac{g^\prime g_X v^2}{2} + M_X M_Y & - \frac{ g g_X v^2  }{2}
	& \frac{g_H g_X \lee v^2 - v^2_\Phi \rii }{2} & g_X^2 \left( v^2 + v^2_\Phi \right) + M_X^2
\end{pmatrix} \; ,
\end{equation}
where $M_X$ and $M_Y$ stand for two Stueckelberg
mass parameters~\cite{Stueckelberg:1938zz,Ruegg:2003ps,Kors:2005uz,Kors:2004iz,Kors:2004ri, Kors:2004dx,Feldman:2007nf,Feldman:2007wj,Feldman:2006wb} associated with $U(1)_X$ and $U(1)_Y$,
respectively. The determinant of this matrix is zero. As a
consequence, there is at least one massless eigenstate which can be
associated with the photon. We set $M_Y = 0$ to avoid a non-zero
electric charges of the neutrinos as discussed in~\cite{Huang:2019obt}. Moreover, this particular choice allows us to bring the
matrix in Eq.~\eqref{M1sq2} into the following form
 \begin{align}
{\cal M}^2_Z = 
\begin{pmatrix}
0 & 0 & 0 & 0\\
0 &
M_{Z^\text{SM}}^2 &
- \frac{g_H v }{2} M_{Z^\text{SM}} &
- g_X v M_{Z^\text{SM}} \\
0 &
- \frac{g_H v}{2} M_{Z^\text{SM}} &
\frac{g_H^{2} \left(v^{2} + v_\Phi^{2}\right)}{4} &
\frac{g_X g_H \left(v^{2} - v_\Phi^{2}\right)}{2}\\
0 &
- g_X v M_{Z^\text{SM}} &
\frac{g_X g_H \left(v^{2} - v_\Phi^{2}\right)}{2} &
g_X^{2} (v^{2} + v_\Phi^{2}) + M_X^{2}
\end{pmatrix} \, 
\label{eq:MgaugeSMrot}
\end{align}
where $M_{Z^{\rm SM}} = \sqrt{g^2 + g^{\prime 2}}\, v/2$ is the SM
gauge boson $Z^{\rm SM}$ mass. This is done by applying Weinberg rotation using the following matrix to the Eq.~\eqref{M1sq2}
\begin{equation}
\label{eq:rot4by4dec}
{\cal O}^{W} =
\begin{pmatrix}
c_W & -s_W & 0 & 0 \\
s_W & c_W & 0 & 0 \\
0 & 0 & 1 & 0 \\
0 & 0 & 0 & 1
\end{pmatrix},
\end{equation}
i.e., $({\cal O}^{W})^{\rm T}\cdot \mathcal{M}_1^2 (M_Y=0)\cdot{\cal O}^{W} $. As a result of this rotation, we get the massless photon associated with the first component of
the basis of this matrix while the second component gives us
the $Z^{\rm SM}$. In this case, we have the intermediate basis
$\mathcal{V}'_Z=\{A,Z^{SM}, W^{\prime 3}, X\}^{\rm T}$. Thus, one can express the
original rotation matrix as $\mathcal{O}^{G}_{4
\times 4}(M_Y=0)
= \mathcal{O}^W \cdot \mathcal{O}^Z$, where the matrix $\mathcal{O}^Z$ brings
$\mathcal{M}^2_Z$ in Eq.~\eqref{eq:MgaugeSMrot} into a diagonal form. 
Consequently, the relation between the mass eigenstates and the intermediates states is given by $\mathcal{V}'_Z = \mathcal{O}^Z\cdot \{A, Z, Z',
Z''\}^{\rm T}$.
For the rest of the paper, we assign $\mathcal{O}^G$ to the non-diagonal $3\times 3$ part of $\mathcal{O}^Z$,
such that $\mathcal{O}^Z_{j+1,k+1} = \mathcal{O}^G_{j,k}$ with $j$ and $k=1,2,3$, as explicitly
given in Eq.~(6) of Ref.~\cite{Huang:2019obt}.
As a remark, the photon $A$ stays the same between
the intermediate states $\mathcal{V}'_Z$ and the mass eigenstates. This
implies that the only non-vanishing component in the first row and column of $\mathcal{O}^Z$ is $\mathcal{O}^Z_{1,1} = 1$.

\subsection{Dark Matter Candidate}
\label{ssec:HiddenP}

As we mentioned in the preceding sections, the stability of the DM
in G2HDM is protected by accidental $\mathcal Z_{2}$ symmetry. This symmetry, also known as the hidden parity ($h$-parity), is respected by all sector in G2HDM even after the SSB~\cite{Huang:2015wts,Chen:2019pnt}. In addition, this parity also forbids FCNC~\cite{Glashow:1976nt,Paschos:1976ay} to occur at tree level for the SM in this model. All particles in G2HDM can be classified according to this parity as shown in Table~\ref{tab:Z2Eff}.   
\begin{table}[htbp!]
	\begin{tabular}{|c|c|c|}
		\hline
		Fields & $h$-parity\\
		\hline
		 $h$, $G^{\pm,0}$, $\phi_{2}$, $G^0_H$, $\delta_{3}$, $f$, $W^{\mu}_{1,2,3}$, 
		 $B_{\mu}$, $X^{\mu}$, $W^{\mu\prime}_{3}$, $G^{\mu a}$   & 1 \\
		 \hline
		 $G^{p,m}_H$, $H_2^0$, $H_2^{0 *}$, $H^{\pm}$, $\Delta_{p,m}$, $f^{H}$, 
		 $W^{\mu \prime}_{1,2}$ & $-1$ \\
		\hline 
	\end{tabular}
	\caption{Classification of all the fields in G2HDM under $h$-parity.} 
	\label{tab:Z2Eff}
\end{table}

In principle, all particles with zero electric charge and have 
$h$-parity equal to -1 are suitable DM candidates. Accordingly, from
Table~\ref{tab:Z2Eff}, the heavy neutrinos $\nu^H$, neutral
gauge boson $W^{\prime (p,m)}$, and the physical complex scalar $D$
satisfy these requirements. In this work, we focus on complex scalar
$D$ as our DM candidate. The mass mixing matrix in Eq.~\eqref{eq:Z2oddmassmatrix} shows that the DM candidate mass
eigenstate, $D$, is linear combination of the interaction
eigenstates $G^p_{H}$, $H^{0*}_{2}$, and $\Delta_{p}$.
Mathematically, this linear superposition is given by 
\begin{equation}
\label{eq:Dcomposition}
D = \mathcal{O}^D_{12} G^p_H + \mathcal{O}^D_{22} H^{0*}_2+ \mathcal{O}^D_{32}
\Delta_p \;,
\end{equation}
where $\mathcal{O}^D_{ij}$ is the $(i,j)$-th component of the rotation matrix $\mathcal{O}^D$. The values of these components depend on the numerical values of the parameters in
Eq.~\eqref{eq:Z2oddmassmatrix}.

Following Ref~\cite{Chen:2019pnt}, we classify our complex scalar DM
candidate into three distinct cases: inert doublet-like DM, triplet-like DM, and Goldstone boson-like DM. This classification depends on
the dominant component of the gauge eigenstate which can be
characterized by the magnitude of $\mathcal{O}^D_{ij}$ in Eq.~\eqref{eq:Dcomposition}. We have inert doublet-like DM if $f_{H_2} \equiv (\mathcal{O}^D_{22})^2 > 2/3$. Triplet-like DM is
achieved if $f_{\Delta_p} \equiv (\mathcal{O}^D_{32})^2 > 2/3$.
Goldstone-like DM is characterized by $f_{G^p} \equiv (\mathcal{O}^D_{12})^2 > 2/3$.  
The magnitude of the $\mathcal{O}^D_{ij}$ elements in~Eq.\eqref{eq:Dcomposition} satisfy the following relation $f_{G^p}+f_{H_2}+f_{\Delta_p}=1$.

In addition, there is another condition that holds for the
Goldstone-like DM. In this case, the value of $f_{G^p}$ must satisfy
$0.67 < f_{G^p} < 0.8$ which is derived from both EWPT constraint
and non-tachyonic DM mass solution as discussed in~\cite{Chen:2019pnt}. Here, we focus on triplet-like DM and Goldstone-like DM.
The Inert doublet-like DM has been ruled out since it can not
survive the relic density and direct detection constraints ~\cite{Chen:2019pnt}.

\section{Dark Matter Experimental Constraints}
\label{sec:constraints}

To understand the properties of DM, we evaluate the DM-SM
interactions via the existing experimental results. These
include the observed DM relic abundance and the limit set by the null result coming from DM
direct search. In this paper, we focus only on these two constraints as they provide the most stringent limit on the complex
scalar DM phenomenology based on the previous study in~\cite{Chen:2019pnt}. Here, we discuss the general feature of these experimental
limits used in this work. 

\subsection{Relic Density}
\label{sec:relic}

Since the interaction between DM and SM is very weak in most of the
models, the observed relic abundance will be typically large.
However, there exist some mechanisms other than DM self annihilations
which enable us to reach the observed abundance.

First, thanks to the common ${\cal Z}_2$-odd
quantum number, DM coannihilates with other ${\cal Z}_2$-odd
particles if their mass splitting is small (typically $\lesssim 10\%
$) such that their number densities do not suffer the Boltzmann
suppression. In this work, coannihilations occur mainly between $D$
and
new heavy fermions $f^{H}$. This happens because of $5\%$ mass
splitting that we impose
on heavy fermions. Furthermore, as we will see later, this
coannihilation is dominated by heavy fermions annihilation,
especially heavy quarks-anti quarks pair ($q^{H}\bar{q}^{H}$) annihilate into $q\bar{q}$ and $gg$. Here, $q\bar{q}$ and $gg$ denote SM quarks and gluon, respectively.

Second, the presence of SM Higgs and heavy Higgs $h_{2}$ resonances significantly increase the thermally-averaged DM annihilation
cross section. In G2HDM, this mechanism allows us to reach the correct relic abundance. 

We use the most recent result
provided by the PLANCK collaboration~\cite{Aghanim:2018eyx} for the relic density,
$\Omega h^2 = 0.120\pm 0.001$, to restrict our thermal relic calculation. Moreover, we set an additional condition on the
parameter space 
of G2HDM to reproduce this result within $2\sigma$ significance.

\subsection{Direct Detection}

The most up to date limit for DM direct search  is provided by
the XENON1T collaboration~\cite{Aprile:2018dbl}.  The zero signal 
result from this experiment strongly restricts the DM nucleon cross section, 
in particular, in the mass regime
between 10~GeV to 100~GeV. Furthermore, for DM mass around 25~GeV, they ruled out the DM-nucleon
elastic cross sections which have the value larger than $10^{-46} \text{~cm}^{2}$.

In G2HDM, it has been shown that the interaction
between DM and nucleon exhibits isospin violation (ISV)~\cite{Chen:2019pnt}. In this case, the ratio between DM-neutron to DM-proton
effective coupling, $f_n/f_p$, is not equal to 1. This occurs due to
the $Z_{i}$ bosons mediated interaction. As an example, for SM $Z$ exchange, the vectorial coupling between quark $q$ ($d$ or $u$-type) and $Z$ is given by~\cite{Chen:2019pnt} 
\bea
\label{eq:qqsz}
g^{V}_{\bar{q} q Z} = \frac{i}{2} \left[ \frac{g}{c_{W}} \left(T_{3} - 2 Q_q s^{2}_{W} \right) \mathcal{O}^{G}_{11} + g_{H} T^{\prime}_{3} \mathcal{O}^{G}_{21} + g_{X} X \mathcal{O}^{G}_{31} \right] \, .
\eea
Here, $Q_q$, $T_3$, $T'_3$ and $X$ stand for the electric charge, the third generator of $SU(2)_{L}$, the third generator of
$SU(2)_{L}$, and the generator of $U(1)_{X}$, respectively. Since
the $u$ and $d$ quark have distinct quantum number assignments with
respect to the underlying gauge group, they couple to the SM $Z$ differently.

In order to accommodate the ISV, we need to calculate the DM-nucleus elastic scattering
cross section $\sigma_{D \mathcal N}$
\begin{equation}
\sigma_{D\mathcal{N}}=\frac{4\mu^2_{\mathcal{A}}}{\pi}
\left[ 
f_p \mathcal{Z} + f_n (\mathcal{A} - \mathcal{Z})
\right]^2 \, ,
\label{eq:sigmaTH}
\end{equation}
where $\mathcal N$ denotes a nucleus with mass number $\mathcal{A}$ and proton number $\mathcal{Z}$ and $\mu_{\mathcal{A}}= m_D m_\mathcal{A}/(m_D +m_\mathcal{A})$ is the reduced mass for DM-nucleus system. 
For definiteness, we neglect all the isotopes of xenon and fix $\mathcal A$ and $\mathcal{Z}$
to 131 and 54, respectively. 
To extract the DM-nucleon effective couplings $f_n$ and $f_p$, we employ 
\texttt{micrOMEGAs}~\cite{Belanger:2018mqt} in our computation.
To compare against the limit given by XENON1T which assumes  $f_n=f_p$ in their analysis, we need to reconstruct their result at the nucleus level.
In order to do so, we follow the same procedure in~\cite{Chen:2019pnt} and obtain the following expression for general value of  $f_n/f_p$ 
\begin{equation}
\sigma^{\rm X1T}_{D\mathcal{N}}=\sigsip({\rm X1T})\times  
\frac{\mu^2_{\mathcal{A}}}{\mu^2_{p}} \times \left[ \mathcal{Z} + \frac{f_n}{f_p} 
\left({\mathcal{A} - \mathcal{Z}} \right) \right]^2 \, ,
\label{eq:sigmaEXP}
\end{equation} 
where $\mu^2_{p}$ is the DM-proton reduced mass. 
In this paper, we use Eq.~\eqref{eq:sigmaEXP} to constrain our direct detection
prediction.

\begin{figure}[!phtb]
	\centering{\includegraphics[height=10cm,width=12cm]{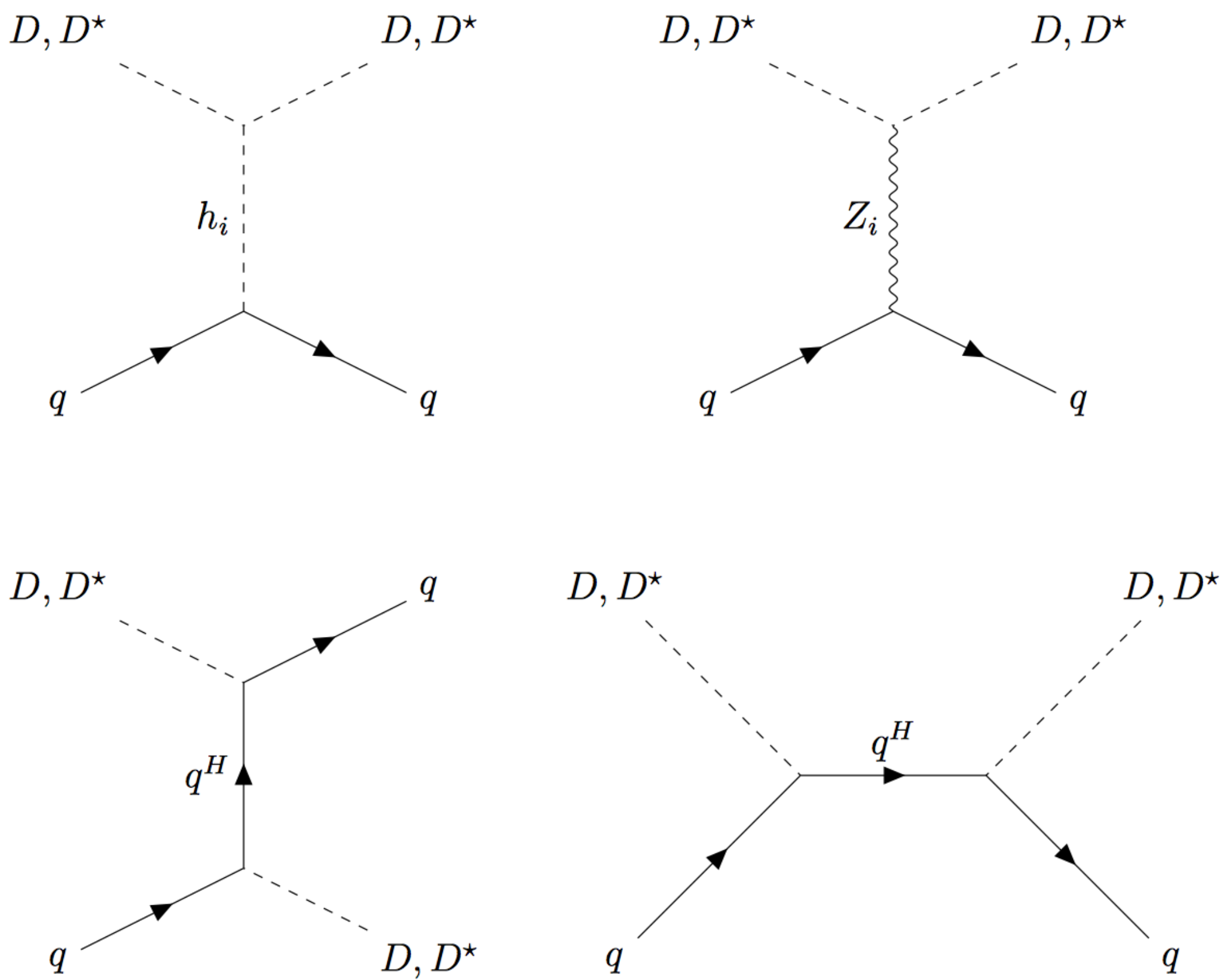}}
	\caption{The Feynman diagrams relevant for DM direct search. They consist of three t-channel diagrams with
the $\mathcal Z_2$-even Higgs bosons (top left),
neutral gauge bosons (top right) and heavy quarks (bottom left)
mediators. There is also s-channel heavy quarks exchange (bottom right). 
}
\label{fig:DDFeynman}
\end{figure}

In general, the DM-nucleon interaction will be different from antiDM-nucleon interaction. The spin independent interaction for complex scalar DM can be obtained by using the effective operator language as~\cite{Belanger:2008sj}
\begin{eqnarray}
\label{eq:scalar_si_lgrgn}
  \mathcal{L}_{D} &=& 2 \lambda_{N,e} M_D D D^*  \bar{\psi}_N\psi_N
  + i \lambda_{N,o} \left( D^* \overset{\longleftrightarrow}{\partial_\mu} D \right)
\bar{\psi}_N\gamma^\mu \psi_N \, ,
\end{eqnarray}
where the $\psi_N$, $\lambda_{N,e}$, and $\lambda_{N,o}$ stand for the nucleon
field operator, the coupling of even operator, and the coupling of odd
operator, respectively. In this case, the  effective coupling between DM (antiDM) and the nucleon is  
\begin{eqnarray}
  \lambda_N =\frac{\lambda_{N,e} \pm
  \lambda_{N,o}}{2} \; ,
  \label{eq:odd-even}
\end{eqnarray}
where the plus (minus) sign denotes DM-nucleon (antiDM-nucleon) interaction. 
Under the interchange between $D$ and $D^{*}$, the first (second)
term in the right hand side of Eq.~\eqref{eq:scalar_si_lgrgn} stays
the same (flips sign). Because of this, the first term is called  \emph{even} operator while the second term describes \emph{odd}
operator.  
As a consequence, the strength of the interaction between DM-nucleon
and antiDM-nucleon is different and it is given by
Eq.~\eqref{eq:odd-even}. 
Hence, the value of $\sigma_{D^* \cal N}$ is in general not equal to 
$\sigma_{D \cal N}$ given by Eq.~\eqref{eq:sigmaTH} because the effective
couplings $f_p$ and $f_n$ for $D$ are different from those for $D^*$. In this work, we average the contribution from both $\sigma_{D \cal N}$ and $\sigma_{D^* \cal N}$ when comparing our calculation with the XENON1T result given in Eq.~\eqref{eq:sigmaEXP}.

The Feynman diagrams relevant for
describing DM-quark
interactions in G2HDM are shown in Fig.~\ref{fig:DDFeynman}.  For the upper part, the left
panel is the $t$-channel interaction mediated by three Higgs bosons while the right panel
is the $t$-channel with three neutral gauge bosons exchange. In addition,
heavy quarks exchange via s and t-channel are shown in the lower
part of the same figure.  
There is a destructive interference between heavy
quarks diagrams and the Higgs mediated diagrams in triplet-like DM ($\Delta_{p}$). On the other hand,
the diagrams with the neutral gauge bosons and heavy quarks mediators add up constructively in Goldstone boson-like DM ($G^{p}_{H}$).

\section{Numerical Analysis and Results}
\label{sec:result}
\subsection{Methodology}

Taking into account the various constraints based on previous G2HDM
studies, we collect a sample of points by doing random scans.
These points pass the scalar sector constraints studied in~\cite{Arhrib:2018sbz} as we mentioned in the preceding section. 
In addition, we also impose the limits from the gauge sector carried out in~\cite{Huang:2019obt}. This study includes
Drell-Yan constraint which restricts the $SU(2)_{H}$ gauge
coupling, $g_{H}$, below 0.1. To prevent the decay of DM into $W^{\prime}$,
we impose the condition  $m_{W^\prime}$ $>$  $m_{D}$ which translates into the lower bound of $g_{H}$~\cite{Chen:2019pnt} 
\begin{equation}
g_{H\text{min}} = \frac{2 m_D}{\sqrt{v^2 + v_\Phi^2 + 4v_\Delta^2}} \, ,
\label{eq:ghmincondition}
\end{equation}
where we have used Eq.~\eqref{eq:Wppmmass}.
The search of $Z'$ enforces us to scan $v_\Phi$ in the mass range between 20 TeV to 100 TeV~\cite{Huang:2019obt}. Moreover, we fix $M_{X} = 2$~TeV to satisfy heavy $M_{X}$ scenario while maintaining
the SM-like $Z$ to $91.1876\pm0.0021$~GeV within 3$\sigma$ significance~\cite{Huang:2019obt}.

In this work, we set the lower limit of heavy fermions  masses to be
no less than
1.0~TeV based on the signal search of the G2HDM new gauge bosons
performed in \cite{Huang:2017bto}.
In previous work~\cite{Chen:2019pnt}, the lower
limit is 1.5~TeV taking from the search of SUSY colored
particles~\cite{lipi:2017}.
Also, we set the new Yukawa couplings,
which are related to the new heavy fermion masses by~\footnote{In
this study, the identity matrix is used to account for the mixing
between different flavors of SM and heavy fermions with the $\mathcal Z_2$-odd scalars in the new Yukawa sectors.}
$m_{f^H} = y_{f^H} v_\Phi/\sqrt 2$, to be 
small enough in order to suppress their contributions on
perturbative unitarity and 
renormalization group running effects. In addition, the $5\%$ mass
splitting between the DM $D$ and heavy fermions $f^{H}$ is imposed
here. This
allows us to study their effects on DM phenomenology which is the
main focus of this paper. In contrast, the $20\%$ mass splitting was
imposed in~\cite{Chen:2019pnt}. 
To accommodate all of these requirements, we
adopt the Yukawa couplings for
each point in our random scan as follows
\begin{equation}
\label{eq:HFYukawa}
y_{f^H} = \text{max}\left[ \frac{1.0\text{ TeV}}{v_\Phi / \sqrt{2}},
\text{min}\left( \frac{1.05\,m_D}{v_\Phi / \sqrt{2}},\; 1 \right) \right]\;.
\end{equation}
From this formula, it is easy to see that the value of $y_{f^H}$ is always less than 1 for all parameters space.
Based on this setup, heavy fermions are expected to give dominant contribution to the DM coannihilation.

Approximately, 3 million points are collected from our random scan.
These points cover all numerical
values for model parameters, gauge bosons and scalar
masses,
as well as all components of three mixing matrices $\mathcal{O}$, $\mathcal{O}^D$ 
and $\mathcal{O}^G$. Furthermore, we input these numbers to
\texttt{MicrOMEGAs}~\cite{Belanger:2018mqt} to compute DM relic density and
DM-nucleon cross section.

The scan range for our computation is tabulated in Table~\ref{tab:scanranges}. Note that 
$M_{H\Delta}$, $M_{\Phi\Delta}$, $v_\Delta$ and $v_\Phi$ have  distinctive ranges for the associated DM composition.
These differences are remarkable in the case of $v_\Delta$ and $v_\Phi$ in
the Goldstone-like column which exhibits a fine-tuning. This happens
because the Goldstone-like composition is achievable only when
$0.8 \leq v_\Delta/v_\Phi < 0.9$ as discussed in~\cite{Chen:2019pnt}.

\begin{table}
\begin{tabular}{|c|c|c|}
\hline
Parameter &  Triplet-like & Goldstone-like \\
\hline \hline
$\lambda_H$                                     & [0.12, 2.75]                      & [0.12, 2.75]      \\
$\lambda_\Phi$                             & [$10^{-4}$, 4.25]                 & [$10^{-4}$, 4.25] \\
$\lambda_\Delta$                            & [$10^{-4}$, 5.2]                  & [$10^{-4}$, 5.2]  \\
$\lambda_{H \Phi}$                             & [$-$6.2, 4.3]                     & [$-$6.2, 4.3]     \\
$\lambda_{H \Delta}$                          & [$-$4.0, 10.5]                    & [$-$4.0, 10.5]     \\
$\lambda_{\Phi \Delta}$                         & [$-$5.5, 15.0]                  & [$-$5.5, 15.0]  \\
$\lambda_{H \Phi}^\prime$                       & [$-$1.0, 18.0]                    & [$-$1.0, 18.0] \\
$\lambda_H^\prime$         & [$-8\sqrt{2}\pi$, $8\sqrt{2}\pi$] & [$-8\sqrt{2}\pi$, $8\sqrt{2}\pi$] \\
\hline
$g_H$                                       & [See text, 0.1]                   & [See text, 0.1]   \\
$g_X$                                       & [$10^{-8}$, 1.0]                  & [$10^{-8}$, 1.0]  \\
\hline
$M_{H\Delta}$/GeV                        & [-3000, 5000] & [0.0, 5000]      \\
$M_{\Phi \Delta}$/GeV                       & [$-$50.0, 50.0] & [0.0, 700]     \\
$v_{\Delta}$/TeV                                 & [0.5, 20.0] & [14.0, 20.0]    \\
$v_{\Phi}$/TeV                                     & [20, 100] & [20, 28.0]      \\
\hline
\end{tabular}
\caption{Parameter ranges used in the numerical scans that pass \textbf{SGSC} constraints. Here, $M_X$ is fixed at 2 TeV and $M_Y=0$.
}
\label{tab:scanranges}
\end{table}

\subsection{Results}
\label{analysis}

Here, we discuss the analysis of our numerical results. We evaluate the effects of heavy fermions on DM phenomenology for triplet-like DM and Goldstone-like DM separately. 

\subsubsection*{$SU(2)_H$ Triplet-like DM}

\begin{figure}[t]
\centering
\includegraphics[width=0.48\textwidth]{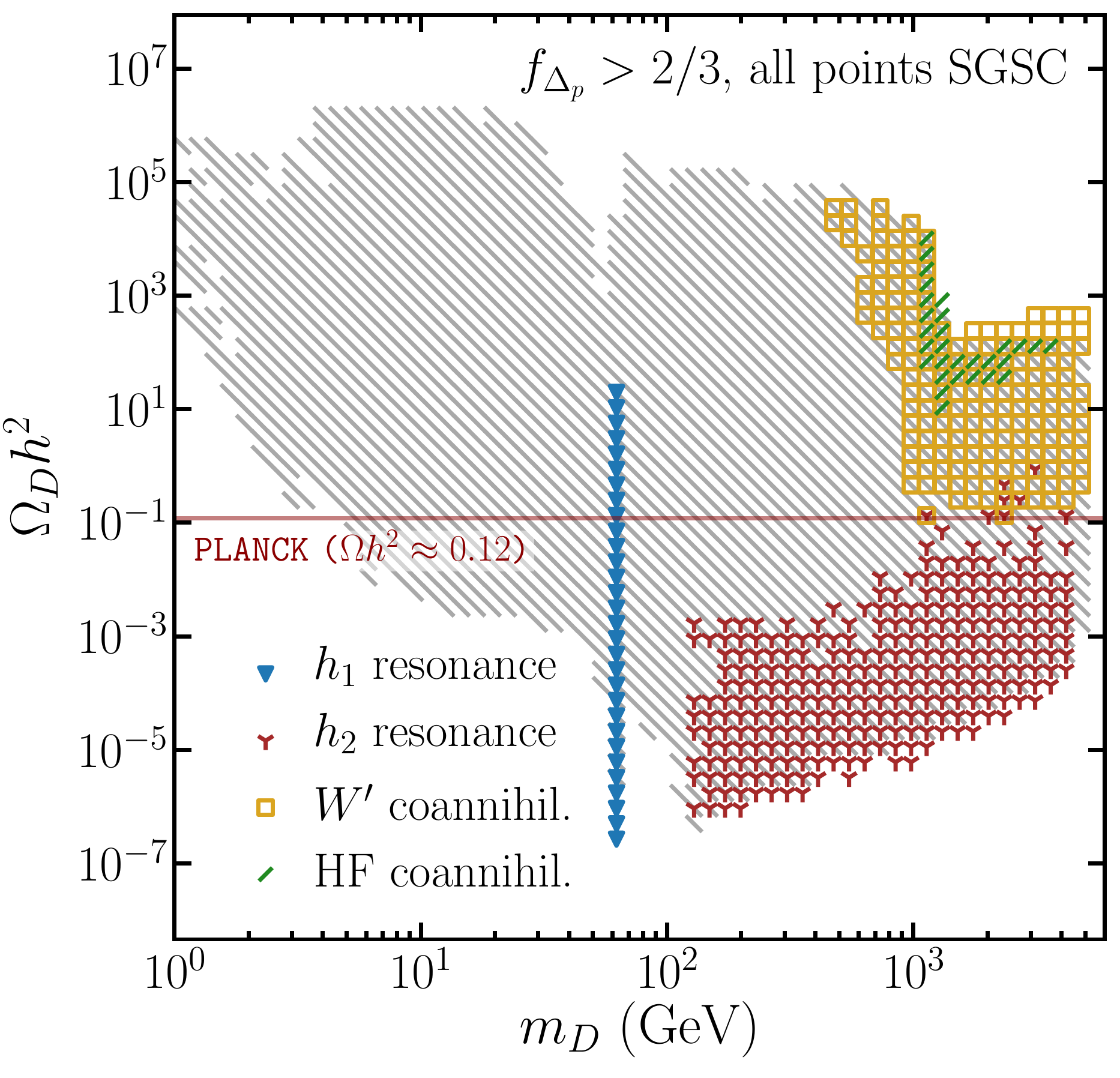}
\includegraphics[width=0.49\textwidth]{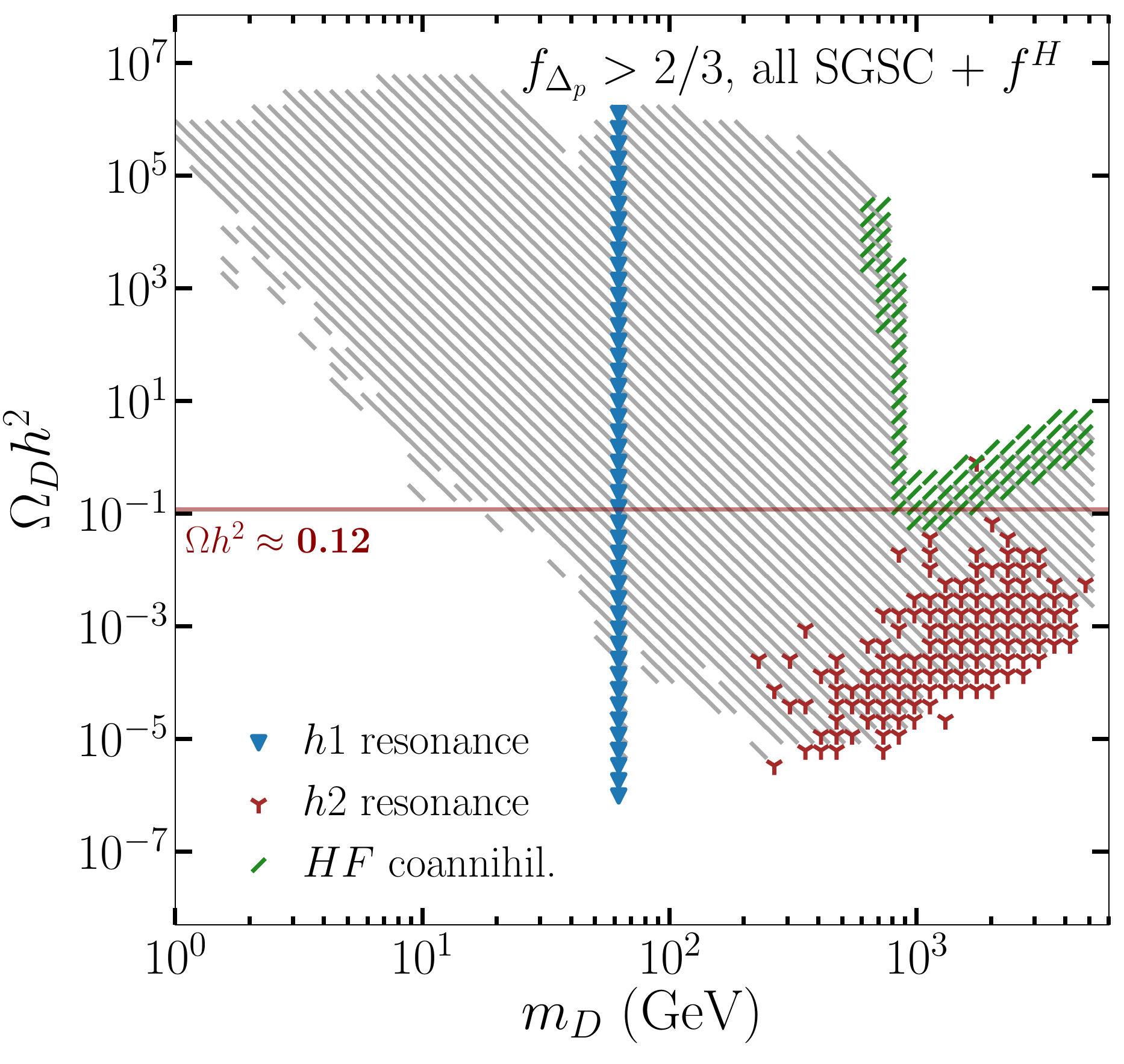}
\caption{The left panel describes triplet-like DM {\bf SGSC} surviving regions projected on ($m_D$, $\oh$) 
adopted from\cite{Chen:2019pnt}. This figure neglects heavy fermions
contributions. The inclusion of heavy fermions (with $5\%$ mass splittig) is depicted on the
right panel. For both cases, the gray area is
ruled out by PLANCK measurement at $2\sigma$.} 
\label{fig:triplet}
\end{figure}

The scatter plots for the relic density versus DM mass $m_D$ in
the triplet-like DM are presented in Fig.~\ref{fig:triplet}.
The red line represents the observed relic density provided by
the PLANCK collaboration.
The right panel takes heavy fermions contributions into account
while the left is adopted from~\cite{Chen:2019pnt}. 
Let us elaborate more by considering different DM mass regimes.

\begin{itemize}
\item[(i)]
We start with the lower DM mass regime ranging from $1\gev$ to $62.5 \gev$. In this region, the DM annihilation cross section is dominated by the following processes: $DD^*\to\
c \bar{c}$, $\tau^+ \tau^-$ and $b \bar{b}$, occuring mainly  via the $s$-channel of the Higgs bosons exchange. This can be understood based on the
$DD^*h_i$ coupling in 
Eq.~\eqref{eq:g_T_ddh}. Thanks to the large value
of $v_\Phi$, the $h_2$ ($\delta_3$-like) mediated diagram is equally as important as the $h_1$ (SM-like) exchange while the heaviest Higgs $h_{3}$ ($\phi_2$-like)
remains subdominant~\cite{Chen:2019pnt}. The observed relic
abundance can be achieved in this wide range ($12\gev < m_{D} < 62 \gev$) due to the various combinations for the $DD^*h_i$ coupling in 
Eq.~\eqref{eq:g_T_ddh}. However, this picture is only valid when
heavy fermions are neglected (the left panel of Fig.~\ref{fig:triplet}). Once they are included (the right panel of Fig.~\ref{fig:triplet}), the t-channel of heavy
fermions exchange becomes relevant. There is a cancellation between
the Higgs and new heavy fermions mediated diagrams 
which reduces the value of DM annihilation cross section. This
enhances the relic density as shown in the right panel of Fig.~\ref{fig:triplet} where the gray points are shifted above the
horizontal red line. The observed relic density is achieved in the mass range between $54\gev$ to $62\gev$.  

\item[(ii)]
When DM mass approaches half of the SM Higgs mass ($m_{D} \approx 62.5 \gev$), the interactions involving the s-channel of $h_{1}$ exchange become dominant. 
This enhances the annihilation cross section resulting in the
suppression of the relic density. However, due to the wide variation
of the $DD^*h_1$ coupling in Eq.~\eqref{eq:g_T_ddh}, the observed
relic abundance can still be achieved. This is also true for $h_{2}$
resonance. In this case, the resonance appears in a broad range of
DM mass thanks to the arbitrary values of $m_{h_{2}}$ ($m_{h_{2}} > m_{h_{1}}$). Note that there is no SM $Z$ resonance because its
coupling is suppressed by $(\mathcal{O}^{D}_{32})^{2} \mathcal{O}^{G}_{21}$ as can be seen from Eq.~\eqref{eq:g_T_ddz}.

\item[(iii)]
Above the SM Higgs resonance, $m_{h_1}/2<m_D<500\gev$, the
major contributions to the DM annihilation cross section are given by  $W^{+}W^{-}$ (more than $\sim 50 \%$),
$h_{1}h_{1}$ ($\sim 25\%$), and $ZZ$ ($\sim 20\%$) final states~\cite{Chen:2019pnt}. In all of these final states, the main
contributions come from the $h_{1}$ and $h_{2}$ exchanges which
lead to the $S-$wave cross section. On the other hand, the $Z_{i}$
mediated diagrams relevant for $W^{+}W^{-}$ final states are $P-$ wave suppressed. 

\item[(iv)]
Next, for heavy DM mass regime ($m_D>500\gev$), the dominant
final states are
given by the longitudinal parts of the SM gauge bosons, $W^{+}_{L}W^{-}_{L}$ and 
$Z_{L}Z_{L}$. However, this is only valid if one neglects heavy fermions contribution. This picture changes in the mass regime larger than 1 TeV in the presence of heavy fermions.

\item[(v)]
Finally, in the TeV region ($m_{D} > 1 \,\text{TeV}$), heavy fermions coannihilations start to dominate. This can
be seen from the right panel of Fig.~\ref{fig:triplet}. The main
coannihilation channels in this regime are given by heavy quarks
annihilations into a pair of SM quark-anti quark and gluons: $q^{H}\bar{q}^{H} \rightarrow q\bar{q},gg$. Based on the points that we collect from
our scan, these processes contribute around 87$\%$ in this regime.
This is expected since these are the QCD interactions controlled by the strong coupling $\alpha_{S}$.
The main diagrams associated with these interactions are given by
the s-channel of gluon exchange for $q\bar{q}$ final states and
the s-channel of gluon exchange, t- and u-channel of $q^{H}$
exchange
for $gg$ final states.   
\end{itemize}

\begin{figure}[!htb]
\includegraphics[width=0.48\textwidth]{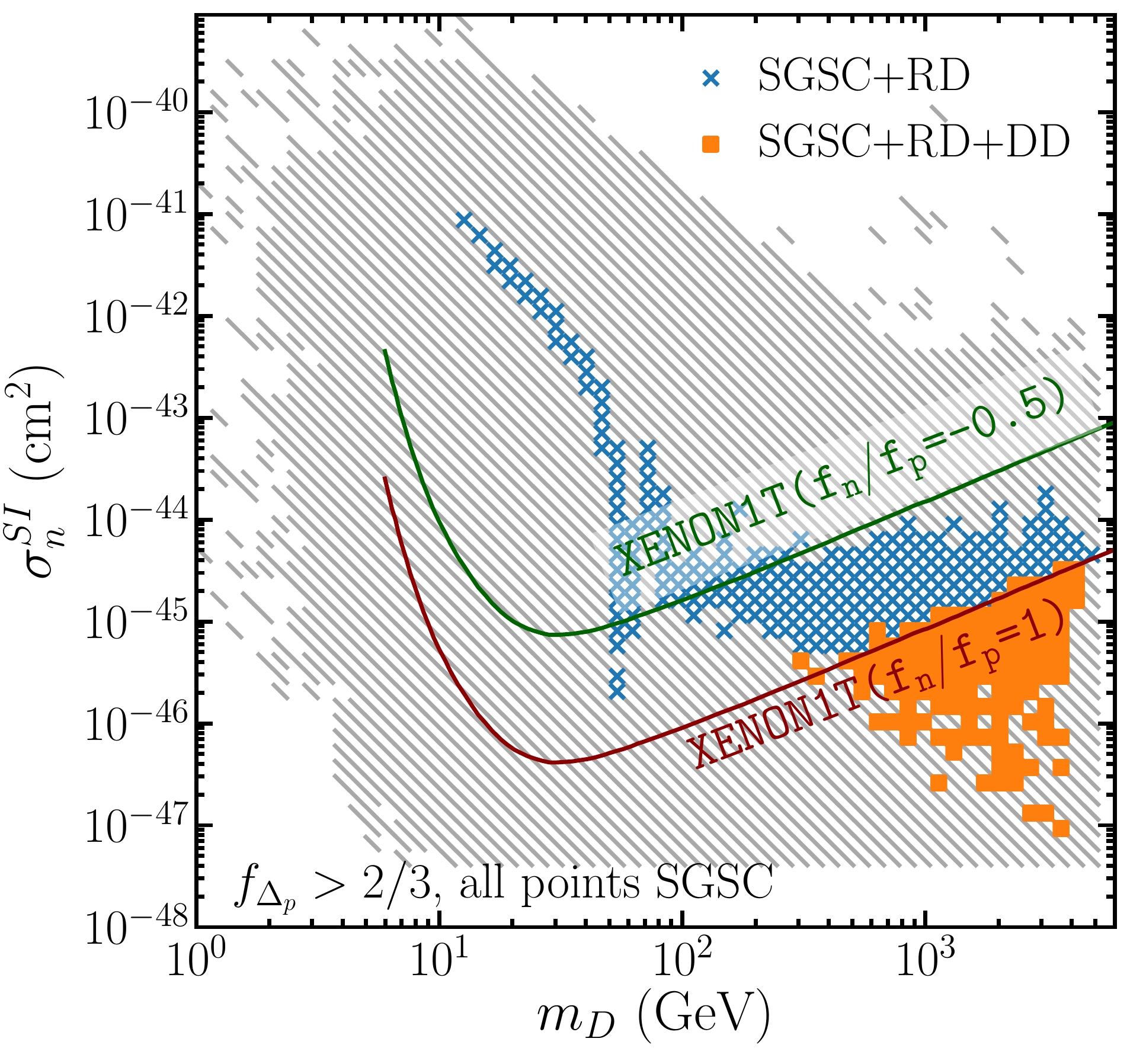}
\includegraphics[width=0.48\textwidth]{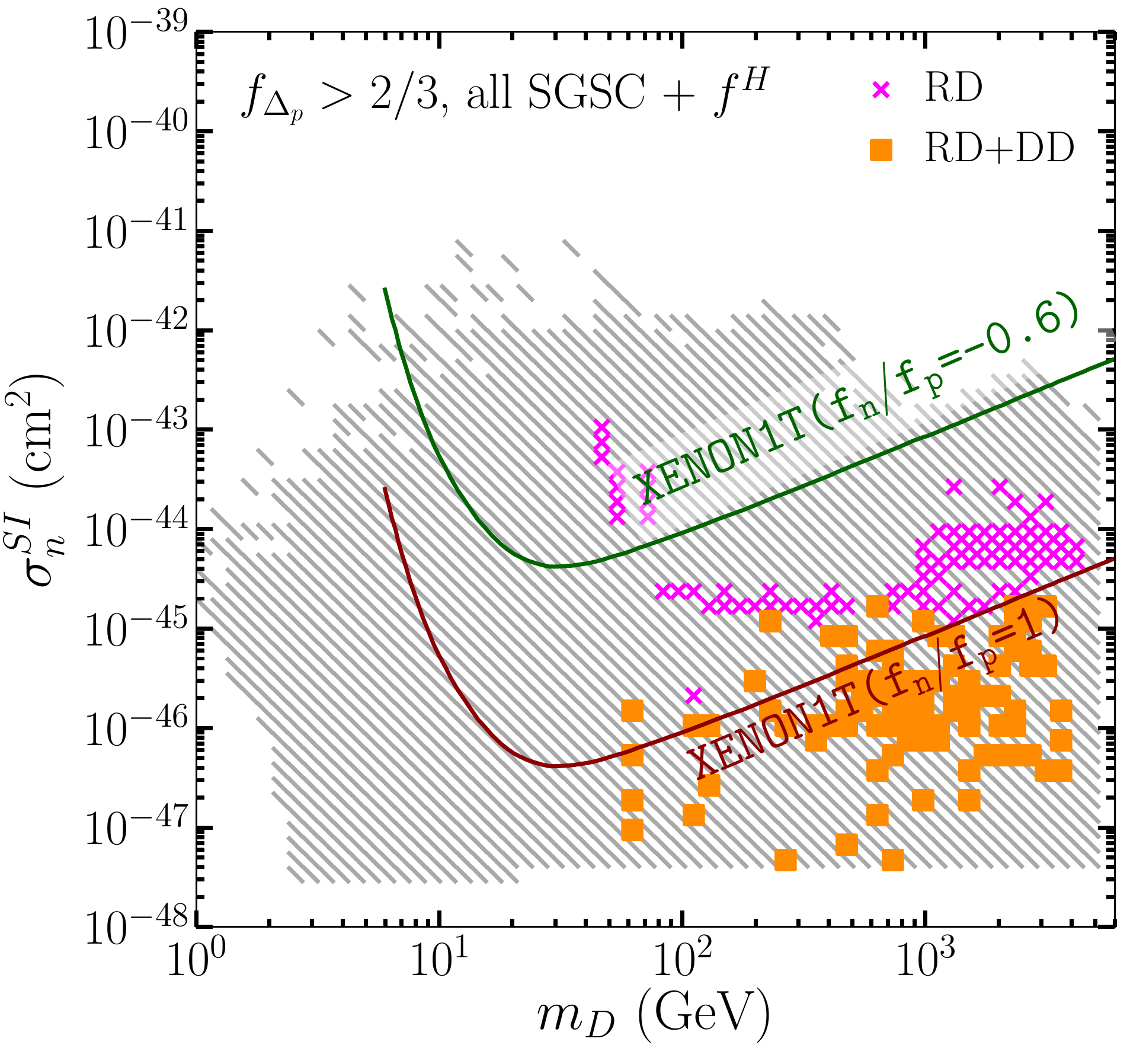}
\caption{In the right panel, adopted from~\cite{Chen:2019pnt} (without heavy fermions), the red solid curve is the official XENON1T exclusion limit without ISV,
while the green solid curve describes the same limit with ISV value $f_n/f_p=-0.5$.
Heavy fermions contribution is shown in the right panel. Here, the green solid line is XENON1T limit with $f_n/f_p=-0.6$. For both figures,
the orange filled squares pass both \textbf{RD} and \textbf{DD} constraints.
\label{fig:TLDD}
}
\end{figure}

In Ref.~\cite{Chen:2019pnt}, the relevant diagrams
for the spin independent DM-nucleon interactions are given by
the t-channel of $h_{i}$ and $Z_{i}$ exchange as displayed in the
upper part of Fig.~\ref{fig:DDFeynman}. The leading contribution comes from $h_{1}$ mediator while $h_{2}, Z$, and $Z'$ mediators
give non-negligible effects. 
This is summarized in the left panel of Fig.~\ref{fig:TLDD}, adopted
from~\cite{Chen:2019pnt}. The XENON1T limit is described by the
solid red curve which assumes $f_{n}/f_{p} = 1$. The orange points in this figure are
allowed by the relic density and direct detection constraints. As
one can see, for DM mass larger than $300\gev$, both
constraints can be evaded. As a result of ISV originating from $Z$
and $Z'$ exchange, there are some orange points which are located slightly above the XENON1T exclusion limit. As a comparison,
the corresponding XENON1T limit for $f_{n}/f_{p} = -0.5$ is also shown.

On the other hand, when heavy fermions contribution is considered, one needs to include all diagrams in
Fig.~\ref{fig:DDFeynman}.
There is a destructive interference between heavy quarks and the Higgs mediated diagrams.
This cancellation can be extracted from the sum of these diagrams as
\begin{align}
\label{eq:cancelHHF}
\mathcal{A}_{(q^{H} + h_{i})} \propto \left[ \frac{m_{q^{H}}\, (\mathcal{O}^{D}_{12})^{2}}{v^{2}_{\Phi}} + \frac{m_{q}\, \mathcal{O}_{1i}\,\lambda_{DD^{*}h_{i}}}{v\, m^{2}_{h_{i}}} \right] \, ,
\end{align}
where $m_{q}$ stands for the mass of the SM quarks. The index $i$
runs from 1 to 2 denoting the
contributions from the Higgs $h_{1}$ and $h_{2}$ while the coupling $\lambda_{DD^{*}h_{i}}$ is given by Eq.~\eqref{eq:g_T_ddh}. The
first term on the right hand side of Eq.~\eqref{eq:cancelHHF} is
always positive while the second two terms, for both $h_{1}$ and $h_{2}$, are mostly negative.

This effect can be seen from the right panel of
Fig.~\ref{fig:TLDD} where most of the gray points are located below  $10^{-42} \text{cm}^{2}$. Interestingly, this cancellation shifts
the allowed DM mass (orange squares area) towards the lighter region, reaching up to $54\gev$. There are some gaps between $60\gev$ and $100\gev$ as well as $140\gev$ and $200\gev$ where none
of the points survive both relic density and direct detection
constraints. This is
because the cancellation given by Eq.~\eqref{eq:cancelHHF} does not
work effectively in this regime. In addition, the existence of the
orange points significantly above the XENON1T limit shows the
importance of the $Z$ and $Z'$ exchange when this cancellation happens efficiently.
 
We see that for triplet-like DM, the
inclusion of heavy fermions put a stronger constraint on the
observed relic abundance, especially in the TeV regime. In
contrast, they reduce the DM-nucleon spin independent cross section
alleviating the current limit from XENON1T experiment.   


\subsubsection*{$SU(2)_H$ Goldstone boson-like DM}
\label{sec:Goldstone}

\begin{figure}[htb]
\centering
\includegraphics[width=0.48\textwidth]{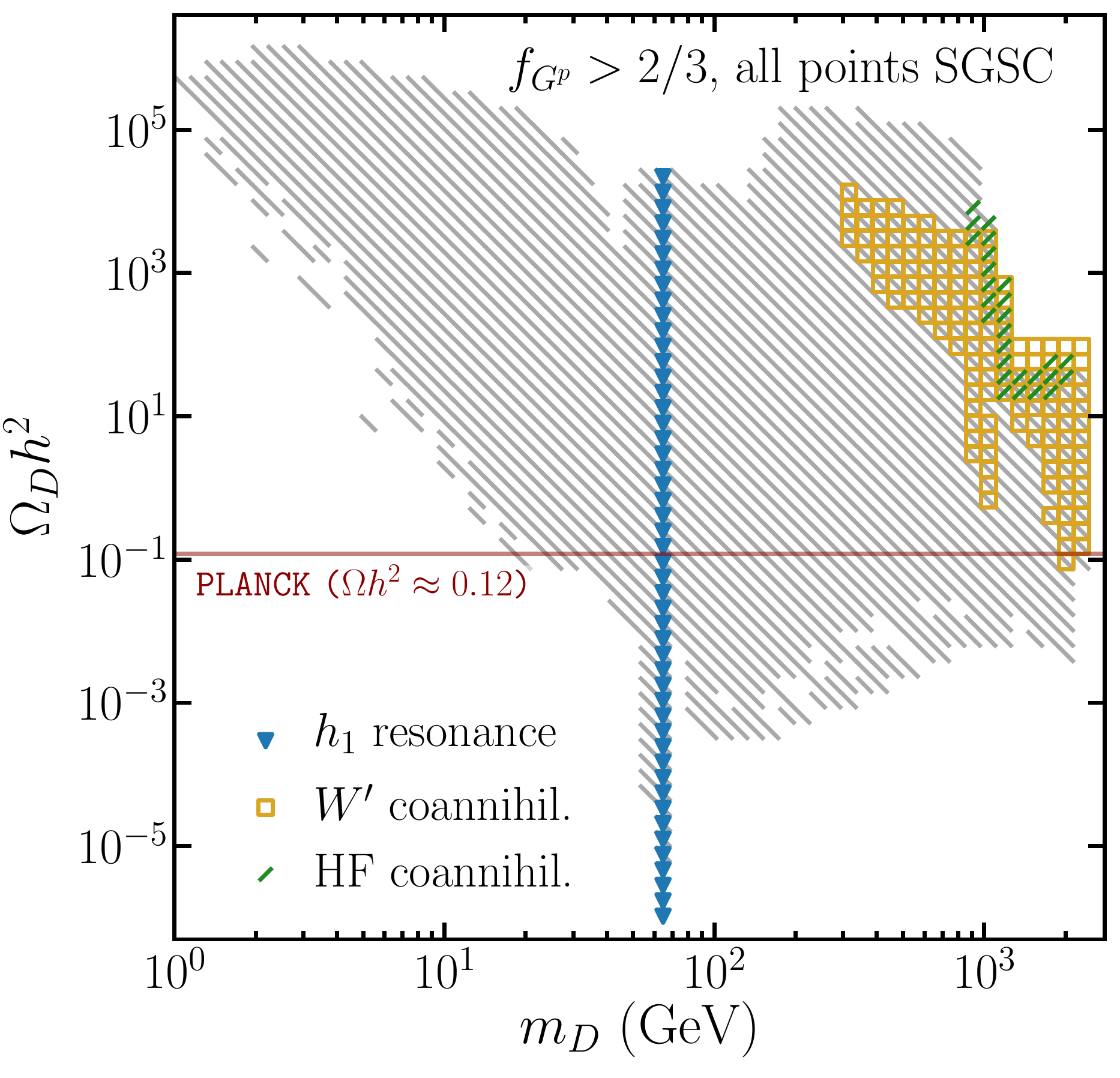}
\includegraphics[width=0.49\textwidth]{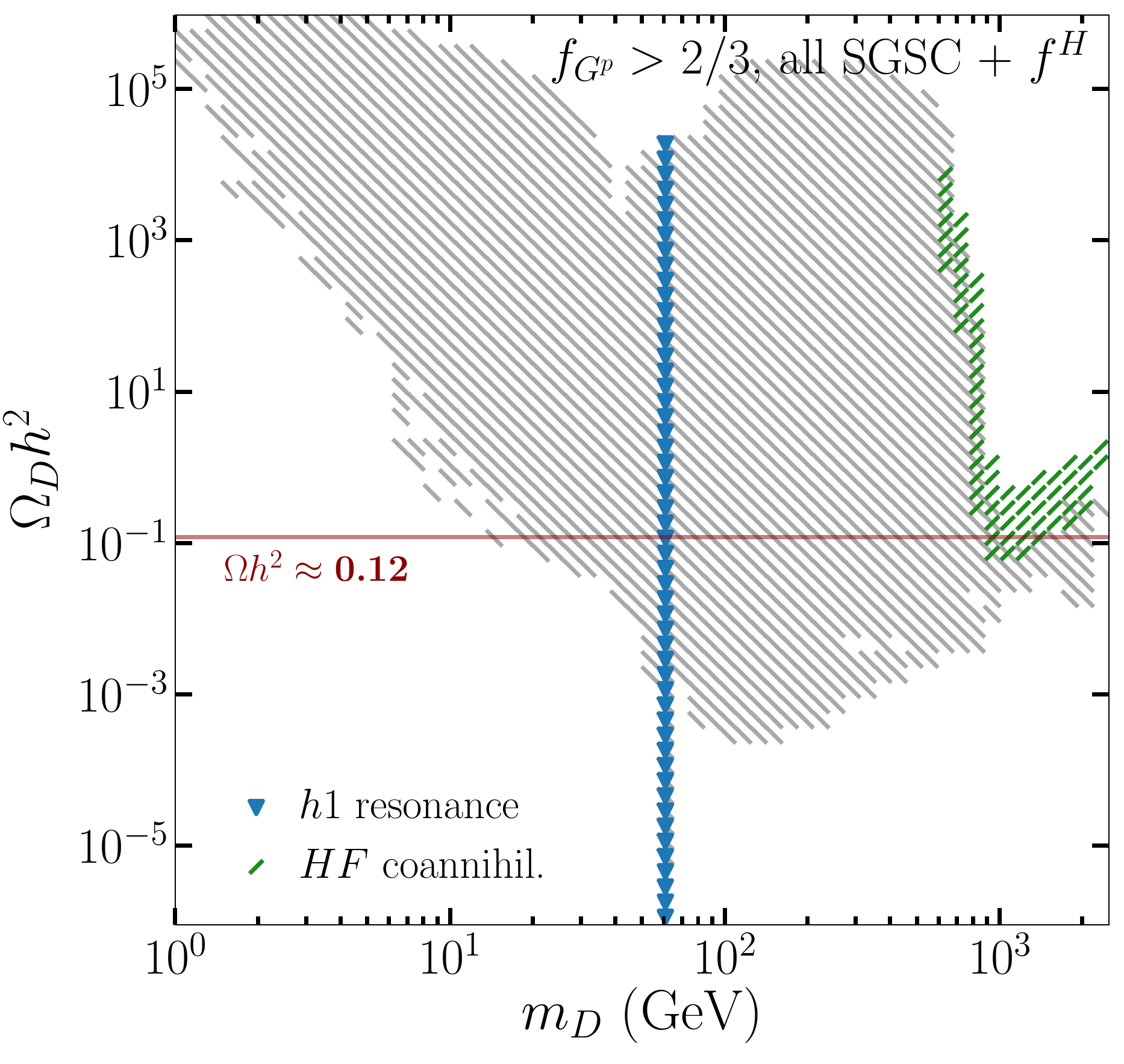}
\caption{The left panel shows Goldstone-like DM {\bf SGSC} allowed regions projected on ($m_D$, $\oh$). This figure, adopted from Ref.~\cite{Chen:2019pnt}, neglects heavy fermions contributions. In contrast, the figure in the right panel takes heavy fermions into account which reduce the relic density in the TeV regime.}
\label{fig:goldstone}
\end{figure}

As we mentioned before, the Goldstone-like DM is dominated by $G^p_H$.
However, in this particular case, the allowed composition lies in a
very narrow regime between $0.67 < f_{G^{p}} < 0.8$. As a result,
this DM lives in a tiny region in the parameter space of G2HDM model.
In the left panel of Fig.~\ref{fig:goldstone}, we show the scatter
plot of the DM relic abundance with respect to its mass. The
underlying physics in each mass regime (regions (i)-(iv)) is mostly
similar to the triplet case~\cite{Chen:2019pnt}. Moreover, when
heavy fermions are included, as shown in the right panel of Fig.~\ref{fig:goldstone}, they strongly dominate the coannihilation
channels in TeV regime as in triplet-like DM case (region (v)).
Furthermore, there is no cancellation between Higgs and heavy fermions exchange thanks to the limited parameter space.      
\begin{figure}[tb]
\includegraphics[width=0.48\textwidth]{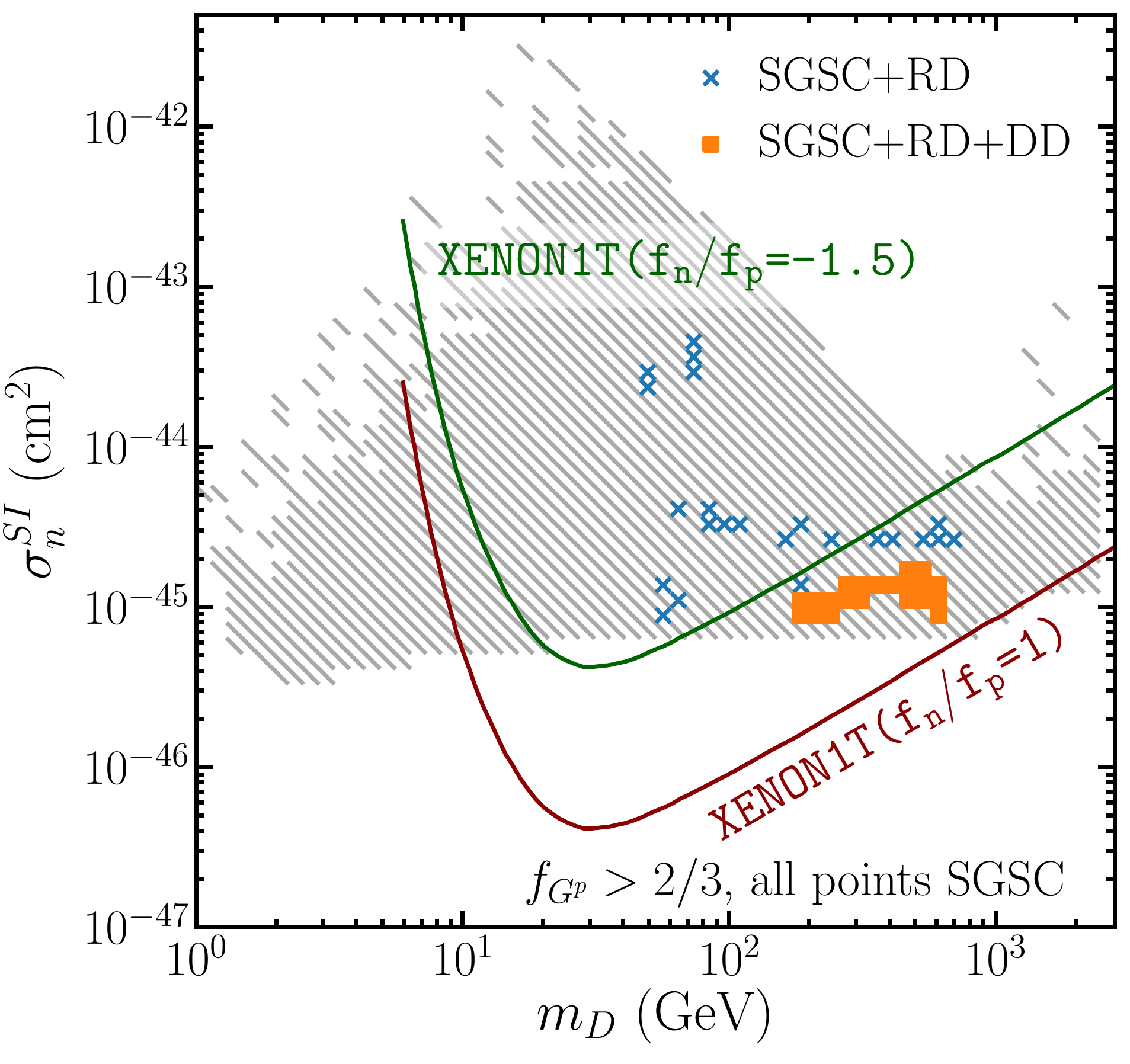}
\includegraphics[width=0.48\textwidth]{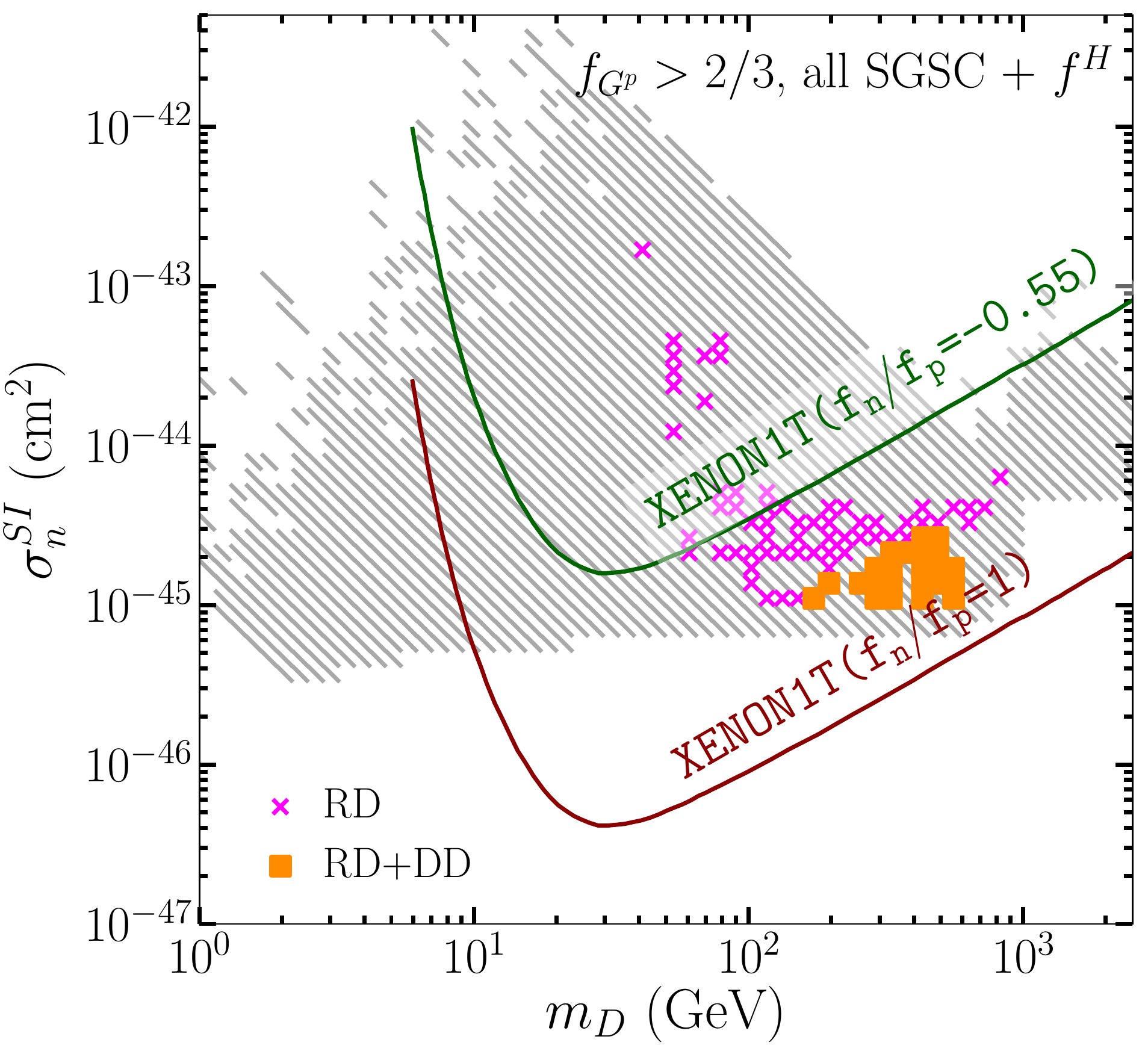}
\caption{The left panel is taken from~\cite{Chen:2019pnt} which omits
heavy fermions. They are included in the figure on the right panel.
The
red solid curve, in both figures,  is the XENON1T limit without ISV.
The green solid line in the left (right) describes the XENON1T
exclusion line with ISV value $f_n/f_p=-1.5$ ($f_n/f_p=-0.55$). 
The small region of orange squares shows the fine-tuned region with $f_n/f_p=-1.86$ for the left panel and $f_n/f_p=-2.00$ for the
right panel. These areas are allowed by \textbf{RD} and \textbf{DD} constraints.
\label{fig:GLDD}
}
\end{figure}


The scatter plot for DM-neutron cross section versus DM mass in the
absence of heavy fermions is
presented in the left panel of Fig.~\ref{fig:GLDD}. The leading
diagram is given by the t-channel of $h_{1}$ exchange which is
significant in the mass range between $3\gev$ and $100\gev$. 
The next important diagrams come from $Z'$ and $Z$ exchange~\cite{Chen:2019pnt}. These ISV diagrams are relevant for DM mass
larger than $100\gev$.  
As a result of the interference between these three dominant
diagrams, the spin
independent cross section varies in a wide range.
However, only the orange points in the mass range $200 \text{~GeV} \leq m_{D} \leq 600 \text{~GeV}$ satisfy both
constraints from the observed relic density and the published
XENON1T limit.
These points have $f_n/f_p = -1.86$ thanks to the fine-tuned
parameter
space. As shown in the right panel of Fig.~\ref{fig:GLDD}, the
inclusion of heavy fermions does not
significantly change this result. They enhance the $Z'$ and $Z$
mediated diagrams shifting the ISV value of the orange region to have $f_n/f_p = -2.0$. This enhancement can also be seen in the TeV
regime where the gray
points are shifted upward.

\section{QCD Sommerfeld Correction and Bound State Formation}
\label{sec:SCBSF}
Based on the previous section, we see that the contributions of
heavy fermions are quite significant,
especially in the TeV regime of the DM mass. In this case, for both
triplet-like and Goldstone-like DM, heavy quarks annihilations ($q^{H} \bar{q}^{H} \rightarrow q\bar{q}, gg$) strongly reduce the relic
abundance due to the QCD interactions. Since these processes
occur in the early universe, at energy scale much higher than the
quark-hadron phase transition era, heavy quarks experience a
long range interaction via gluon exchange. This interaction can be described by Coulomb-like potential as~\cite{ElHedri:2017nny}
\begin{align}
\label{eq:GCoulomb}
V(r) = -\frac{\zeta^{'}}{r} , \; \;\;\;\; \; \zeta^{'} = (C_{q^{H}}-\frac{3}{2})\alpha_{s} \, ,
\end{align}
where $\zeta^{'}$ describes the modified coupling between $q^{H}$ and $\bar{q}^{H}$ free pair which depends on
the quadratic Casimir coefficient of the color representation of the
heavy quark $C_{q^{H}}$. The presence of the Coulomb-like potential modifies the perturbative calculation. This effect is known as the Sommerfeld
correction~\cite{Sommerfeld:1931}. For heavy quarks,
the
value of $C_{q^{H}}$ is equal to $4/3$ since they belong to the
color triplet representation of $SU(3)_{C}$.
Furthermore, the presence of this potential may induce the bound
state formation of $q^{H} \bar{q}^{H}$ pairs in the early universe\cite{Feng:2009mn, vonHarling:2014kha, Petraki:2015hla, Kim:2016zyy, Kim:2016kxt, Petraki:2016cnz, Petraki:2014uza, Detmold:2014qqa, An:2016gad, Kouvaris:2016ltf, Ellis:2015vaa, Ellis:2015vna, Nagata:2015hha, Feng:2010zp, Iminniyaz:2010hy, Hryczuk:2010zi, deSimone:2014pda, Keung:2017kot, Liew:2016hqo}.
Therefore, one needs to consider these two effects when calculating
the DM abundance. In this work, we follow the method introduced in~\cite{Liew:2016hqo} for our QCD Sommerfeld correction and QCD bound
state effect calculation.  

\subsection{QCD Sommerfeld Correction}
The thermally averaged $S-$ and $P-$wave annihilation cross sections
of $q^{H} \bar{q}^{H}$ into pair of SM quark and gluons can be expanded as
\begin{align}
\label{eq:spwave}
\left\langle \sigma v_{rel} \right\rangle = a + b \, (T/m_{q^{H}}) + \mathcal{O}((T/m_{q^{H}})^{2}) ,
\end{align} 
where $a$ and $b$ denote the $S-$ and $P-$wave term, respectively. 
For $q\bar{q}$ final state, the corresponding expression is given by\cite{Srednicki:1988ce, Ellis:1999mm, Liew:2016hqo}
\begin{align}
\label{eq:spwaveqq}
\left\langle \sigma v_{rel} \right\rangle_{q\bar{q}} = \frac{\pi\,\alpha^{2}_{s}}{m^{2}_{q^{H}}} \, \left[\frac{4}{3} - \frac{14}{3} \, (T/m_{q^{H}}) \right] ,
\end{align}
while for gluon pair final state, the associated cross section is
\begin{align}
\label{eq:spwavegg}
\left\langle \sigma v_{rel} \right\rangle_{gg} = \frac{\pi\,\alpha^{2}_{s}}{m^{2}_{q^{H}}} \,\left[\frac{7}{27} + \frac{1}{6} \, (T/m_{q^{H}}) \right].
\end{align}
In general, the presence of the Coulomb potential of the form $V(r)=-\alpha/r$ will affect both $S-$ and $P-$wave terms. For the $S-$wave term, it becomes $a \, S(\alpha/v_{rel})$ where 
\begin{align}
\label{eq:Ssommerfeld}
S(\alpha/v_{rel}) = \frac{2\pi\, \alpha / v_{rel}}{1-e^{-2\pi \alpha / v_{rel}}} ,
\end{align}
where $v_{rel}$ stands for the relative velocity between $q^{H}$ and $\bar{q}^{H}$. The Sommerfeld correction enhances the annihilation
cross section if the potential is attractive ($\alpha > 0$) while
the suppression occurs when the potential is repulsive ($\alpha < 0$). Taking this correction into account in the early universe, we need to evaluate the thermally-averaged Sommerfeld $S-$wave cross
section as
\begin{align}
\label{eq:STsommerfeld}
a\, \left\langle S(\alpha/v_{rel}) \right\rangle = a\,\int^{\infty}_{0} S(\alpha/v_{rel})\, f(v_{rel})\, dv_{rel} \, ,
\end{align}
where the velocity distribution $f(v_{rel})$ is given by the Maxwell-Boltzmann distribution
\begin{align}
\label{eq:MB}
f(v_{rel}) = \left( \frac{\mu}{2\pi T}\right)^{3/2} 4\pi v^{2}_{rel}\, e^{-\frac{\mu v^{2}_{rel}}{2T}}\, ,
\end{align}
where $\mu = m_{q^{H}}/2$ and $T$ are the reduced mass of the $q^{H}\bar{q}^{H}$ system and the temperature, respectively. In this work, we only consider the Sommerfeld correction in the $S-$wave term. 

Following two body color decomposition for the $S-$wave cross section, the
thermally averaged Sommerfeld corrections are~\cite{deSimone:2014pda} 
\begin{eqnarray}
\label{eq:Sommqg}
\left\langle \sigma \, v_{rel}(q^{H}\bar{q}^{H} \rightarrow gg) \right\rangle_{s,Somm} &=& \left\langle \sigma v_{rel} \right\rangle_{s,gg} \left( \frac{2}{7}\,\left\langle S(\frac{4\alpha_{s}/3}{v_{rel}})\right\rangle + \frac{5}{7} \left\langle S(\frac{-\alpha_{s}/6}{v_{rel}})\right\rangle \right) \nonumber \\
\left\langle \sigma \, v_{rel}(q^{H}\bar{q}^{H} \rightarrow q\bar{q}) \right\rangle_{s,Somm} &=& \left\langle \sigma v_{rel} \right\rangle_{s,q\bar{q}} \left\langle S(\frac{-\alpha_{s}/6}{v_{rel}})\right\rangle \, ,
\end{eqnarray} 
where $\left\langle \sigma v_{rel} \right\rangle_{s,q\bar{q}}$ and $\left\langle \sigma v_{rel} \right\rangle_{s,gg}$ denote the $S-$ wave part of Eq.~\eqref{eq:spwaveqq} and Eq.~\eqref{eq:spwavegg}, respectively.

\subsection{QCD Bound State Formation}
When $q^{H}$ and $\bar{q}^{H}$ form a bound state $\eta$, the Coulomb-like
potential due to gluon exchange becomes
\begin{eqnarray}
\label{eq:BSFCoulomb}
V(r) &=& -\frac{\zeta}{r} ,\\
\zeta &=& \frac{1}{2}(C_{q^{H}}+C_{\bar{q}^{H}}-C_{\eta})\alpha_{s} \, ,
\end{eqnarray}
where $C_{\eta}$ is the quadratic Casimir operator of the $q^{H}$ $\bar{q}^{H}$ bound state in a particular color state.
Here, we focus on the color singlet bound state where $C_{\eta} = C_{\textbf{1}} = 0$.
Furthermore, we only consider bound states with total angular
momentum $L=0$ and spin $S=0$. The corresponding binding energy of
the bound state $E_{B}$ and its Bohr radius $a$ are
\begin{eqnarray}
\label{eq:EBa}
E_{B} &=& \frac{\zeta^{2}\, \mu}{2} ,\\
a &=& (\zeta\, \mu)^{-1} \, ,
\end{eqnarray}  
where $\mu = m_{q^{H}}/2$ is the reduced mass of the $q^{H}$ and $\bar{q}^{H}$ system.

To calculate the bound state formation cross section, one
needs to know the integrated dissociation cross section first. A $q^{H}$ and $\bar{q}^{H}$ bound state can be destroyed by absorbing
a gluon. The dissociation cross section averaged over the incoming
gluon color is
\begin{eqnarray}
\label{eq:BSD}
\sigma_{dis} &=& \frac{1}{8} \times \frac{4}{3} \times\frac{2^{9}\, \pi^{2}}{3} \alpha_{s} a^{2} \left(\frac{E_{B}}{\omega}\right)^{4} \frac{1+\nu^{2}}{1+(\kappa\,\nu)^{2}} \frac{e^{-4\nu \text{arccot}(\kappa \nu) -2\pi \nu}}{1-e^{-2\pi \nu}} \kappa^{-1}.
\end{eqnarray}
In this expression we define $\nu$, $\kappa$, and $\omega$ as
\begin{align}
\label{eq:nu}
\nu \equiv \frac{|\zeta^{'}|}{v_{rel}} \; \; \; \kappa \equiv \frac{\zeta}{|\zeta^{'}|} \; \; \; \omega \approx E_{B} + \frac{1}{2} \mu v^{2}_{rel} \, .
\end{align}
Eq.~\eqref{eq:BSD} describes the dissociation
cross section of a bound state after absorbing a gluon with
energy $\omega$. The first factor in this equation is the averaged
over incoming gluon color while the second number denotes the color
factor of the
bound state wave function. The bound state formation cross section can be obtained by using the Milne relation as
\begin{align}
\label{eq:BSF}
\sigma_{bsf} = \frac{g_{\eta} \, g_{g}\,\omega^{2}}{g_{q^{H}}\,g_{\bar{q}^{H}}\,(\mu v_{rel})^{2}} \, \sigma_{dis} \, ,
\end{align}
where $g_{g}=16$, $g_{q^{H}}= g_{\bar{q}^{H}}=6$, and $g_{\eta}=1$
are the degree of freedom of gluon, $q^{H}$, $\bar{q}^{H}$, and color singlet bound state, respectively.

In addition, a $q^{H}$ and $\bar{q}^{H}$ bound state can experience
annihilation decay and individual decay of each $q^{H}$ ($\bar{q}^{H}$). In the former case, the $q^{H}$ and $\bar{q}^{H}$
annihilate into the SM particles while in the latter case, each $q^{H}$ ($\bar{q}^{H}$) decays into DM and SM particle. Thanks to the $5\%$
mass splitting between the DM and heavy quarks, the latter process
is suppressed. Additionally, this decay channel is further suppressed by $(\mathcal{O}^{D}_{12})^{2}$. Therefore, in our case, we only need to
consider the annihilation decay channel. The main decay mode is given by two gluon final state~\cite{Kahawala:2011pc}
\begin{align}
\label{eq:qHandecay}
\Gamma^{q^{H}}_{\eta} = \frac{2}{3} \mu \alpha^{2}_{s} \zeta^{3}\,.
\end{align}
Note that in Eq.~\eqref{eq:qHandecay}, the explicit strong coupling
$\alpha_{s}$ is evaluated at the scale of $2m_{q^{H}}$ while the
inverse Bohr radius $a^{-1}$ is used when evaluating the 
$\alpha_{s}$ dependence in $\zeta$.

Considering the DM relic, we have to take the thermal average of the
bound state effect. To get rid of the temperature
dependence when evaluating the thermal average, we define $z\equiv E_{B}/T$ and $u\equiv \frac{1}{2} \mu v^{2}_{rel}/T$ such that we
can express $\sigma_{dis}$ and $\sigma_{bsf}$ in terms
of $z$ and $u$. The bound state dissociation rate after taking the
thermal averaging into account is
\begin{align}
\label{eq:Tdis}
\left\langle \Gamma^{q^{H}}_{\eta} \right\rangle_{dis} = g_{g}\, \frac{4\, \pi}{(2\pi)^{3}} \int^{\infty}_{0} \sigma_{dis}\, \frac{E^{3}_{B}(1+\frac{u}{z})^{2}}{z(e^{z+u}-1)} \, du\, .
\end{align}  
The thermally-averaged bound state formation cross section times velocity is given by
\begin{align}
\label{eq:Tbsf}
\left\langle\sigma v \right\rangle_{bsf} = \int^{\infty}_{0} \sigma_{bsf}\, \zeta \, \left(\frac{u}{z}\right)^{1/2}  \frac{2}{\sqrt{\pi}} u^{1/2}\, e^{-u}\, \left(1 + \frac{1}{e^{z+u} - 1} \right) \, du \, .
\end{align} 
Finally, the thermally-averaged bound state annihilation decay rate is written as
\begin{align}
\label{eq:Tdec}
\left\langle \Gamma^{q^{H}}_{\eta} \right\rangle = \Gamma^{q^{H}}_{\eta}\, \frac{K_{1}(m_{\eta}/T)}{K_{2}(m_{\eta}/T)} \, ,
\end{align} 
where $m_{\eta} = 2m_{q^{H}}-E_{B}$ and $K_{1,2}(m_{\eta}/T)$ are
the bound state mass and the modified Bessel functions of the second kind, respectively.

Taking both Sommerfeld correction as well as bound
state effect into account, we write the thermally-averaged effective annihilation cross section $\left\langle \sigma_{eff} v \right\rangle$ as
\begin{align}
\label{eq:effsv}
\left\langle \sigma_{eff} v \right\rangle =\frac{1}{2} \left\langle \sigma v \right\rangle_{q^{H}\bar{q}^{H}\rightarrow q\bar{q},gg} \frac{g^{2}_{q^{H}}(1+\Delta)^{3}\,e^{-2\Delta x}}{g^{2}_{eff}} \, .
\end{align}   
In this expression, we have defined the following parameters
\begin{eqnarray}
\label{eq:Tdef}
x & \equiv & \frac{m_{D}}{T} \, ,\\
\Delta &\equiv & \frac{m_{q^{H}}-m_{D}}{m_{D}} \, ,\\
g_{eff} & \equiv & g_{D} + g_{q^{H}} (1+\Delta)^{3/2} e^{-\Delta x}\, ,
\end{eqnarray}
where $g_{D} = 6$ is the DM degree of freedom. The explicit expression of $\left\langle \sigma v \right\rangle_{q^{H}\bar{q}^{H}\rightarrow q\bar{q},gg}$ is given by
\begin{eqnarray}
\label{eq:SBSFsv}
\left\langle \sigma v \right\rangle_{q^{H}\bar{q}^{H}\rightarrow q\bar{q},gg} &=& \left\langle \sigma \, v_{rel}(q^{H}\bar{q}^{H} \rightarrow q\bar{q}) \right\rangle_{s,Somm} + \left\langle \sigma \, v_{rel}(q^{H}\bar{q}^{H} \rightarrow gg) \right\rangle_{s,Somm}\nonumber\\
&+& \left\langle \sigma \, v_{rel}(q^{H}\bar{q}^{H} \rightarrow q\bar{q}) \right\rangle_{p} + \left\langle \sigma \, v_{rel}(q^{H}\bar{q}^{H} \rightarrow gg) \right\rangle_{p}\nonumber\\
&+& \frac{1}{4} \left\langle \sigma v_{rel }\right\rangle_{bsf}\, \frac{\left\langle \Gamma^{q^{H}}_{\eta} \right\rangle}{\left\langle \Gamma^{q^{H}}_{\eta} \right\rangle + \frac{1}{4} \left\langle \Gamma^{q^{H}}_{\eta} \right\rangle_{dis}}\, ,
\end{eqnarray}  
where $\left\langle \sigma \, v_{rel}(q^{H}\bar{q}^{H} \rightarrow q\bar{q}) \right\rangle_{p}$  and $\left\langle \sigma \, v_{rel}(q^{H}\bar{q}^{H} \rightarrow gg) \right\rangle_{p}$ stand for the $P-$
wave part of the Eq.~\eqref{eq:spwaveqq} and \eqref{eq:spwavegg}.
The factor of $1/4$ in front of $\left\langle \sigma v_{rel }\right\rangle_{bsf}$ and $\left\langle \Gamma^{q^{H}}_{\eta} \right\rangle_{dis}$ is needed since we only consider the $S=0$ bound state. The DM relic abundance can be calculated by~\cite{Berlin:2014tja}
\begin{align}
\label{eq:relic}
\Omega_{D}h^{2} \cong \frac{x_{f} \times 1.07 \times 10^{9}\gev^{-1}}{g^{1/2}_{*} m_{\text{Pl}} \left\langle \sigma_{eff} v \right\rangle} \, ,
\end{align}
where $x_{f}$, $g_{*}$, and $m_{\text{Pl}}$ are the value of $x$ at
the freeze-out temperature, the number of massless degree of
freedom associated with the energy density, and the Planck mass,
respectively.
In this work, we take all the 6 heavy quarks ($u^{H}$, $d^{H}$, $s^{H}$, $c^{H}$, $b^{H}$, and  $t^{H}$) into account 
when calculating the Sommerfeld correction as well as bound state
effect.
\begin{figure}[tb]
\includegraphics[width=0.98\textwidth]{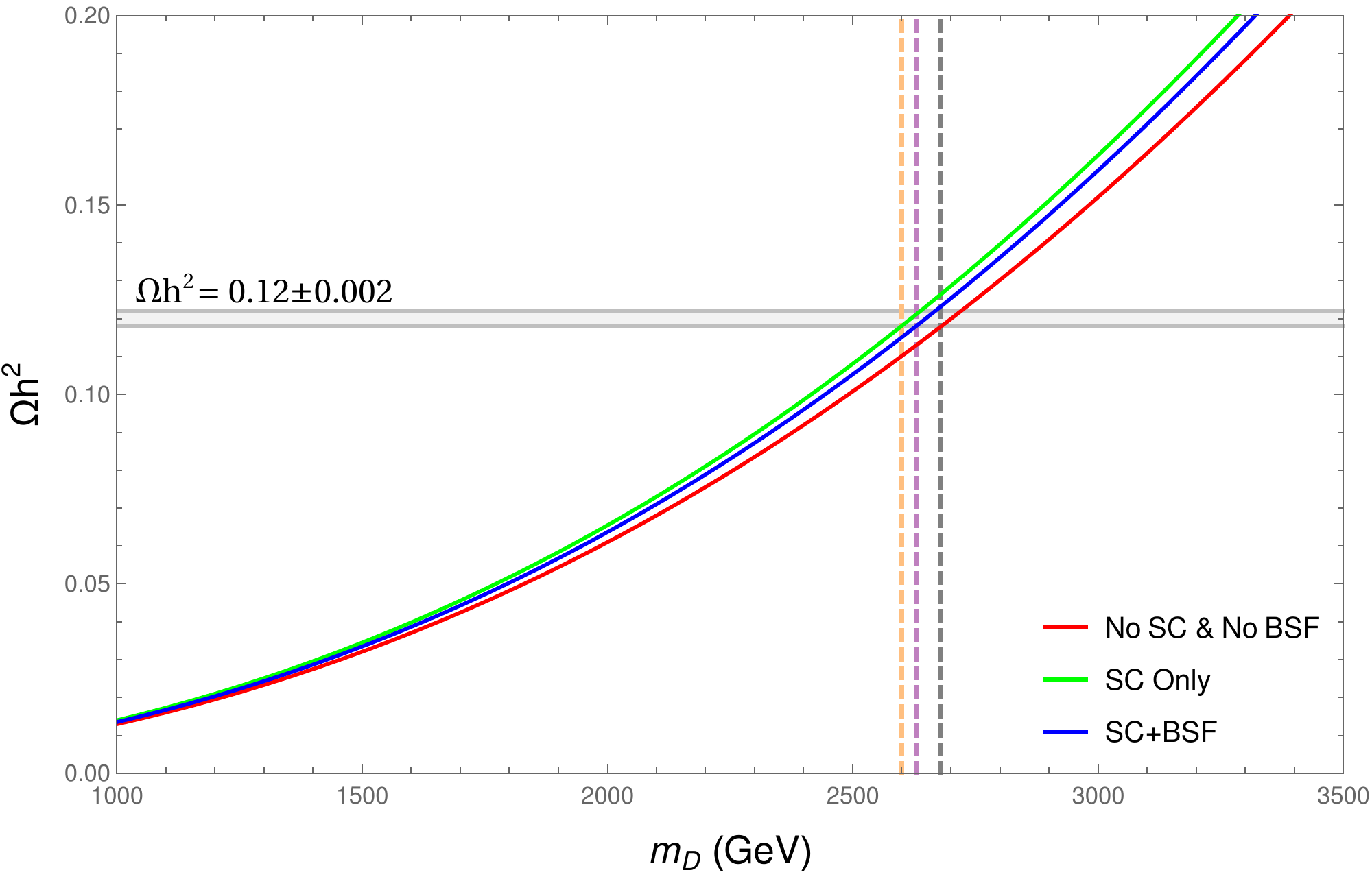}
\caption{DM relic density plotted against its mass for three different cases: perturbative calculation (red solid line) without
Sommerfeld correction (SC) and bound state formation (BSF),
perturbative calculation with SC in $S-$wave (green solid line), and
by considering both SC and BSF (blue solid line). The gray band is
the observed PLANCK data within 2$\sigma$.
\label{fig:SCBSF}
}
\end{figure} 
Furthermore, we only consider one point from our scan to see
how these two effects change our results. This is reasonable because
$q^{H}\bar{q}^{H} \rightarrow q\bar{q},gg$ annihilation cross
sections populate $87\%$ of our scan
points in the TeV regime. 

The effects of QCD Sommerfeld correction and QCD bound state on DM relic
density are shown in Fig.~\ref{fig:SCBSF}. The gray band in this
figure represents the observed DM relic from PLANCK collaboration
within $2\sigma$ significance. The red solid curve comes from the
perturbative calculation given in Eq.~\eqref{eq:spwaveqq} and Eq.~\eqref{eq:spwavegg}. The Sommerfeld correction, as described by the
green solid curve, reduces the perturbative cross section due to the
cancellation between the Sommerfeld enhancement in $q^{H}\bar{q}^{H}
\rightarrow gg$ and the Sommerfeld suppression for $q^{H}\bar{q}^{H}
\rightarrow q\bar{q}$~channels~\cite{ElHedri:2017nny}. As a result,
it increases the DM relic density, shifting the observed relic
towards the smaller DM mass.  

The vertical dashed lines in Fig.~\ref{fig:SCBSF} denote the lower bound
of DM mass that satisfies the observed relic abundance. 
The black dashed line is located at 2680 GeV. This corresponds to
the perturbative calculation. At 2600 GeV, the orange dashed line
comes from the Sommerfeld correction. This $3\%$ shift  
demonstrates that the Sommerfeld correction does not
significantly affect the DM relic abundance. The bound state effect 
along with the Sommerfeld correction is shown by the blue solid
curve. The lower bound of the DM mass is shifted from 2600 GeV to
2630 GeV as given by the
purple dashed line. This shows that the bound state effect, in
contrast to the Sommerfeld effect, slightly reduces the DM relic density.

\begin{figure}[tb]
\includegraphics[width=0.98\textwidth]{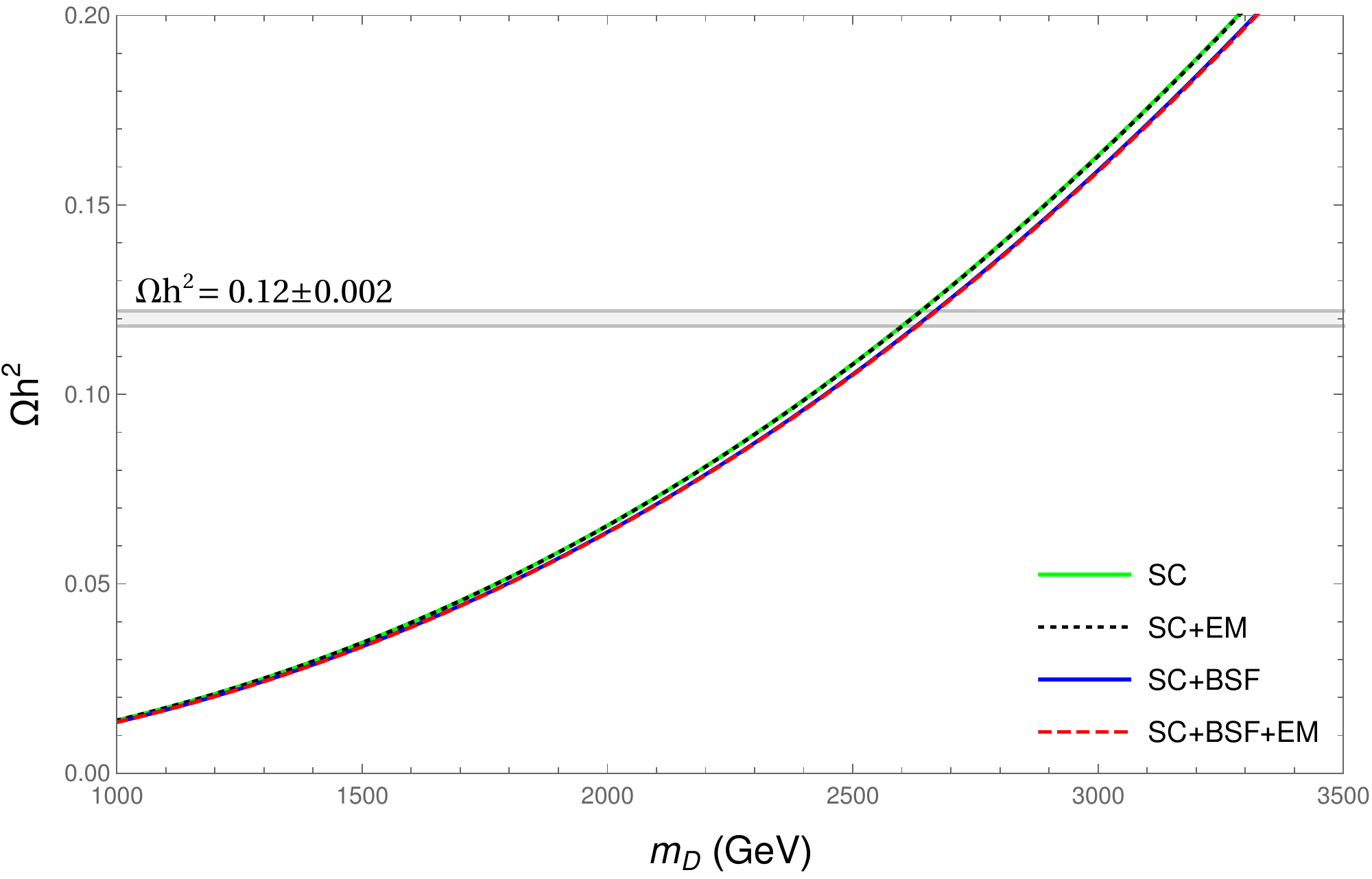}
\caption{The EM correction for Sommerfeld correction (SC) is given by the black dotted line. The red dashed line represents the EM correction for both SC and bound state effect (BSF). The gray band is
the observed relic denisty from PLANCK collaboration within 2$\sigma$ 
\label{fig:SCBSFEM}
}
\end{figure}

When the electric charge of each $q^{H}$ and $\bar{q}^{H}$, $Q_{q^{H}}$, is considered, the Coulomb-like potential  between them becomes more attractive. 
This electromagnetic (EM) correction modifies the coupling of the
corresponding Coulomb-like potentials as $\zeta \rightarrow \zeta +
\alpha_{\text{EM}}\, Q^{2}_{q^{H}}$ and $\zeta^{'} \rightarrow
\zeta^{'} + \alpha_{\text{EM}}\, Q^{2}_{q^{H}}$, where $\alpha_{\text{EM}}$ is the EM fine structure constant. All quantities depending on these couplings are changed as well. The
EM
correction for both Sommerfeld and the bound state effects are 
shown in Fig.~\ref{fig:SCBSFEM}. As in Fig.~\ref{fig:SCBSF}, the
green (blue) solid curve describes the Sommerfeld (Sommerfeld and
the bound state)
effect. The black dotted curve is the EM correction of the
Sommerfeld effect. This weakens the value of the relic density by $0.15\%$ at 2600 GeV. Furthermore, another similar reduction is also
observed when one considers the EM correction of the bound state
effect
along with the Sommerfeld correction.
This is shown by the red dashed curve in Fig.~\ref{fig:SCBSFEM}.
In this case, the reduction of the relic density is about $0.17\%$
at 2630 GeV. These results 
demonstrate that the EM correction is negligibly small and can be
omitted in the calculation of the DM relic abundance.

We see that non-perturbative effects such as the QCD Sommerfeld
correction as well as the QCD bound state effect do not
significantly change the DM relic abundance. Thus, one can safely
ignore these effects and carry out the perturbative calculation
given by Eq.~\eqref{eq:spwaveqq} and Eq.~\eqref{eq:spwavegg}.    

\section{Constraining Parameter Space in G2HDM}
\label{sec:ps}

We see from the preceding sections that the Goldstone-like DM
lives in the limited parameter space region. Moreover, the
particular value 
of $f_n/f_p\approx -2.00$ is needed to avoid the XENON1T limit.
Thus, in this subsection, we focus our discussion on the triplet-like DM when evaluating the surviving parameter space in G2HDM.

\begin{table}[htbp!]
\centering
\begin{tabular}{|c|c|}
\hline 
Allowed Parameter Ranges without $f^{H}$ & Allowed Parameter Ranges with $f^{H}$ \\
\hline 
6.35$ \times 10^{-3} < \lambda_{\Phi} <$ 4.09 & 4.24$ \times 10^{-3} < \lambda_{\Phi} <$  3.98   \\
\hline 
-3.39 $ < \lambda_{H \Delta} <$ 4.07 & -3.16 $  < \lambda_{H \Delta} <$ 4.16   \\
\hline 
-0.07 $ < \lambda_{\Phi \Delta} <$ 6.62 & -0.30 $  < \lambda_{\Phi \Delta} <$ 6.10   \\
\hline 
-5.67 $ < \lambda_{H \Phi} <$ 3.41 & -5.33 $  < \lambda_{H \Phi} <$ 2.54   \\
\hline 
-0.01 $ < \lambda^{'}_{H \Phi} <$ 15.90 & 0.01 $  < \lambda^{'}_{H \Phi} <$ 14.36   \\
\hline 
1.29$ \times 10^{-1} < \lambda_{H} <$ 2.80 & 1.30$ \times 10^{-1} < \lambda_{H} <$  2.69   \\
\hline 
-22.74 $  < \lambda^{'}_{H} <$ 9.57 & -21.75 $  < \lambda^{'}_{H} <$  7.61   \\
\hline 
1.01$ \times 10^{-4} < \lambda_{\Delta} <$ 4.99 & 1.20$ \times 10^{-4} < \lambda_{\Delta} <$  4.84   \\
\hline
7.16$ \times 10^{-3} < g_{H} <$ 0.10 & 3.37$ \times 10^{-3} < g_{H} <$  0.10   \\
\hline
1.01$ \times 10^{-8} < g_{X} <$ 3.55 $\times 10^{-2}$ & 1.01$ \times 10^{-8} < g_{X} <$  3.01 $\times 10^{-2}$   \\
\hline 
\end{tabular}
\caption{The surviving parameter space after DM constraints.} 
\label{tab:Cparam1}
\end{table}

We collect the allowed parameter regions from 
the \textbf{SGSC+RD+DD} constraints
in table ~\ref{tab:Cparam1} and ~\ref{tab:Cparam2}.
In this work, the allowed parameter regions are collected in the
right column of these tables.
The left column lists the surviving parameter space
from~\cite{Chen:2019pnt}. 
As a general remark, we see that 
the inclusion of heavy fermions put the allowed parameter
space into the smaller region. 

As discussed in \cite{Chen:2019pnt}, the \textbf{RD+DD} constraints
put a strong limits on $\lambda_{\Phi\Delta}$, $\lambda_{H\Delta}$,
and $\lambda_{\Phi\Delta}$. This can be seen from their explicit
appearance in $DD^*h_j$ couplings in Eq.~\eqref{eq:g_T_ddh}. Due to
its correlation with $\lambda_{\Phi\Delta}$, these constraints 
restrict $\lambda_{\Phi}$ as well. As a consequence of the loose
requirement that we impose on triplet-like DM, $f_{\Delta_{p}} > 2/3$,
the $G^{p}_{H}$ component becomes non-negligible. In fact, there
are points with 
$G^{p}_{H} = 1/3$ in our scan. Consequently, 
this constraints $\lambda_{H\Phi}$ and
$\lambda'_{H\Phi}$ which are relevant for the $f_{G^{p}_{H}}$ component as one can see
from Eq.~\eqref{eq:g_G_ddh}. Furthermore, there are indirect
constraints on other parameters due to the mixing angles which appear in Eq.~\eqref{eq:g_T_ddh} and Eq.~\eqref{eq:g_G_ddh}.

From previous sections, especially the relic density of triplet-like
DM, we see the importance of the $DD^*h_j$ couplings. In \cite{Chen:2019pnt}, these couplings play important role in the relic density
calculation for all DM mass. However, these couplings are relevant
only in the GeV
regime (region (i) to (iv)) when heavy fermions are included.
This significantly cuts the
allowed parameter space in G2HDM. The same reduction also holds for 
$g_{H}$ and $g_{X}$. These two couplings correspond to the $Z_{i}$
mediated diagrams, especially for $W^{+}W^{-}$ as well as $W^{+}_{L}W^{-}_{L}$
final states in region (iii) and (iv) (although these contributions
are $P-$wave suppressed). Furthermore, $g_{H}$ and $v_{\Phi}$
receive additional constraints from the lower allowed DM mass as 
in Eq.~\eqref{eq:ghmincondition}.
\begin{table}[htbp!]
\centering
\begin{tabular}{|c|c|}
\hline 
Allowed Parameter Ranges without $f^{H}$(in GeV) & Allowed Parameter Ranges with $f^{H}$ (in GeV) \\
\hline 
2.75 $  < M_{H\Delta} <$ 4.99 $\times 10^{3}$ & -2.44$ \times 10^{3} < M_{H\Delta} <$  4.51 $\times 10^{3}$   \\
\hline 
0.01 $  < M_{\Phi \Delta} <$ 49.9  & 0.00 $  < M_{\Phi \Delta} <$ 49.8    \\
\hline 
5.00 $\times 10^{2}  < v_{\Delta} <$ 2.00 $\times 10^{4}$ & 5.07$ \times 10^{2} < v_{\Delta} <$  1.99 $\times 10^{4}$   \\
\hline 
4.15 $\times 10^{4} < v_{\Phi} <$ 1.00 $\times 10^{5}$ & 3.99 $\times 10^{4} < v_{\Phi} <$ 9.99 $\times 10^{4}$  \\
\hline 
\end{tabular}
\caption{The surviving dimensionful parameter space after DM constraints.} 
\label{tab:Cparam2}
\end{table}

Finally, the constraints from DM physics are not sensitive 
to limit $v_{\Delta}$, $M_{\Phi\Delta}$, and $M_{H\Delta}$.
In this paper, we extend the scan range for $M_{H\Delta}$ to
include the negative values of the parameter space. The motivation is to
collect more points in our scan. This explains the broader allowed
parameter range for $M_{H\Delta}$ as one can see from table~\ref{tab:Cparam2}. 


\section{Summary and Conclusion}
\label{sec:summary}

The motivation of this paper is to study the effects of heavy
fermions 
on the complex scalar DM phenomenology which is omitted in the
previous work.
Taking the constraints from the observed relic abundance provided by
PLANCK collaboration
as well as the null result from XENON1T experiment into account,
we compare our calculations against these experimental data. 
By imposing $5\%$ mass splitting between the DM and heavy fermions
as well as lowering their mass to 1 TeV,
we demonstrate that the heavy fermions significantly affect
the complex scalar DM phenomenology.
The DM relic abundance is reduced drastically in the TeV regime
of the DM mass. 
This happens mainly because of the QCD interactions, $q^{H}\bar{q}^{H}\rightarrow q\bar{q}, gg$. 

We further include the QCD Sommerfeld correction as well as 
the QCD bound state effect in our relic density calculation. We
learn that these non-perturbative effects do not change the
perturbative calculation in 
a significant way. On the other hand, heavy fermions weaken the
direct search limit given by the XENON1T experiment. This is due to
their destructive contribution in the relevant Feynman diagrams.
These results put additional constraints on the parameter space of
the G2HDM model. This makes the allowed region in the parameter 
space becomes narrower.
We conclude that the inclusion of heavy fermions on the complex
scalar DM phenomenology is crucial and can not be neglected.


\section*{Acknowledgments}
We would like to thank Prof.~Tzu-Chiang Yuan who encourages the
author to write this paper.
CSN would like to thank Prof. Hsiang-Nan Li for useful discussions
regarding the evaluation of the strong coupling at different scale
in QCD bound state effect. BD would like to acknowledge support from Thailand Science Research and Innovation and Suranaree University of Technology through SUT-Ph.D. Scholarship Program for ASEAN.

\newpage

\appendix

\section{Feynman Rules}
\label{sec:appendix}

Here we collect the most important couplings to the DM
analysis in various interactions as discussed in the text. The gauge couplings $g_H$ and $g_X$ are related to $SU(2)_H$ and $U(1)_X$ gauge group, respectively.
For the scalar-scalar-gauge derivative vertices, 
we use the convention that all momenta are incoming.
\subsection*{Dominant Couplings for Triplet-like DM}

\begin{minipage}{0.3\textwidth}
      \includegraphics{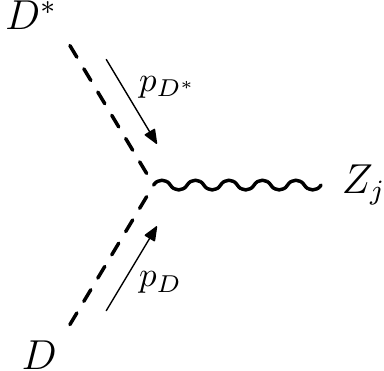}
\end{minipage}
\begin{minipage}{0.69\textwidth}
\begin{flalign}
& \approx i g_{H} (\mathcal{O}^{D}_{32})^{2} \mathcal{O}^{G}_{2j} (p_{D^{*}} -
 p_{D})_{\mu} &
\label{eq:g_T_ddz}
\end{flalign}
\vspace{0.0mm} 
\end{minipage}

\begin{minipage}{0.3\textwidth}
      \includegraphics{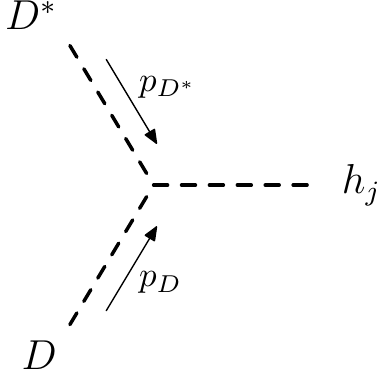}
\end{minipage}
\begin{minipage}{0.69\textwidth}
\begin{flalign}
& \approx i \left[-\lambda_{H \Delta} v  \mathcal{O}_{1j} -
\lambda _{\Phi \Delta } v_{\Phi }  \mathcal{O}_{2j} + 2 \lambda _{\Delta }
v_{\Delta }  \mathcal{O}_{3j} \right](\mathcal{O}^{D}_{32})^{2} &
\label{eq:g_T_ddh}
\end{flalign}
\vspace{0.0mm} 
\end{minipage}

\subsection*{Dominant Couplings for Goldstone boson-like DM}

\begin{minipage}{0.3\textwidth}
      \includegraphics{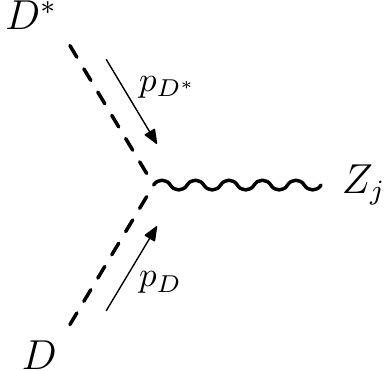}
\end{minipage}
\begin{minipage}{0.69\textwidth}
\begin{flalign}
& \approx i \left[ \frac{g_{H}}{2} \mathcal{O}^{G}_{2j} + g_{X}
\mathcal{O}^{G}_{3j}\right] (\mathcal{O}^{D}_{12})^{2} (p_{D^{*}} -
p_{D})_{\mu} &
\label{eq:g_G_ddz}
\end{flalign}
\vspace{0.0mm} 
\end{minipage}

\begin{minipage}{0.3\textwidth}
      \includegraphics{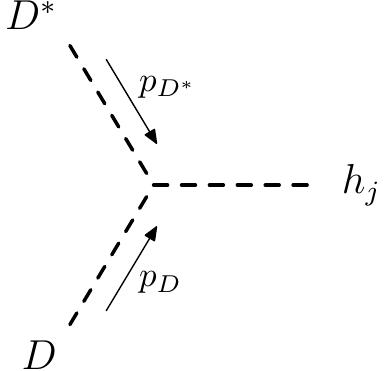}
\end{minipage}
\begin{minipage}{0.69\textwidth}
\begin{flalign}
& \approx i \left[-(\lambda_{H\Phi} + \lambda^{\prime}_{H\Phi}) \mathcal{O}_{1j}  v  -
2\lambda_{\Phi}  \mathcal{O}_{2j}  v_{\Phi} + \lambda_{\Phi \Delta}
\mathcal{O}_{3j}  v_{\Delta} \right](\mathcal{O}^{D}_{12})^{2} &
\label{eq:g_G_ddh}
\end{flalign}
\end{minipage}


\end{document}